\documentclass[aps,prl,reprint]{revtex4-2}

\usepackage{epsfig,amsmath}
\usepackage{balance}
\usepackage{subfigure}
\usepackage{graphicx}
\usepackage{dcolumn}
\usepackage{mathrsfs}
\usepackage{amsthm}
\usepackage{amssymb}
\usepackage{bm}
\usepackage{latexsym}
\usepackage[colorlinks=true,urlcolor=blue,linkcolor=blue,citecolor=blue]{hyperref}
\usepackage{color}
\usepackage{array}
\usepackage{placeins}

\begin{document}
	
\title{Forked Physics Informed Neural Networks for Coupled Systems of Differential equations}

\begin{abstract}
Solving coupled systems of differential equations (DEs) is a central problem across scientific computing. While Physics Informed Neural Networks (PINNs) offer a promising, mesh-free approach, their standard architectures struggle with the multi-objective optimization conflicts and local optima traps inherent in coupled problems. To address the first issue, we propose a Forked PINN (FPINN) framework designed for coupled systems of DEs. FPINN employs a shared base network with independent branches, isolating gradient pathways to stabilize training. We demonstrate the effectiveness of FPINN in simulating non-Markovian open quantum dynamics governed by coupled DEs, where multi-objective conflicts and local optima traps often cause evolutionary stagnation. To overcome this second challenge, we incorporate an evolution regularization loss that guides the model away from trivial solutions and ensures physically meaningful evolution. We demonstrate the effectiveness of FPINN in simulating non-Markovian open quantum dynamics governed by coupled DEs, where multi-objective conflicts and local optima traps often cause evolutionary stagnation. For the spin-boson and XXZ models, FPINN accurately captures hallmark non-Markovian features, such as quantum coherence revival and information backflow, significantly outperforming standard PINNs. The proposed FPINN architecture offers a general and effective framework for solving coupled systems of equations, which arise across a broad spectrum from classical physics to modern artificial intelligence, including applications in multi-body rotational dynamics, multi-asset portfolio optimization, chemical reaction kinetics, and deep representation learning.
\end{abstract}

\author{Zhao-Wei Wang$^{1}$, Zhao-Ming Wang$^{1,2,3}$\footnote{wangzhaoming@ouc.edu.cn}}
\affiliation{
$^{1}$ College of Physics and Optoelectronic Engineering, Ocean University of China, Qingdao 266100, China\\
$^{2}$ Engineering Research Center of Advanced Marine Physical Instruments and Equipment of Ministry of Education, Ocean University of China, Qingdao, China\\
$^{3}$ Qingdao Key Laboratory of Advanced Optoelectronics, Ocean University of China, Qingdao 266100, China}
\maketitle

\textit{Introduction---}
The solution of coupled DEs is a cornerstone of scientific computing, critical to understanding phenomena across fluid dynamics \cite{cai2021physics, MAO2020112789}, electromagnetics \cite{shukla2021parallel}, and non-Markovian quantum dynamics \cite{tanimura2020numerically,bai2024hierarchical,wang2021quantum, diosi1997non, PhysRevLett.83.4909}.
Conventional numerical approaches discretize the governing DEs on a spatiotemporal grid, typically combining spatial discretization methods such as finite difference \cite{courant1928differenzengleichungen} or finite volume \cite{STYNES199583} with Runge–Kutta-based solvers for temporal integration \cite{NIEGEMANN2012364, LIU2024109055}.
However, to address their reliance on mesh discretization \cite{Zerroukat2000Explicit, DIVO2005136} and the curse of dimensionality in high-dimensional problems \cite{DOOSTAN20094332}, these traditional numerical methods have recently been adapted into artificial intelligence algorithms that primarily follow three paradigms.
The first is the time series prediction, a purely data driven approach that learns the mapping of state evolution from large volumes of historical data, often as a black-box model \cite{han2019review, lin2021simulation, wu2021forecasting, JunDong2026Optimal}. 
The second paradigm encompasses neural network based variational methods, which reformulate physical functionals, such as energy, into neural network representations and optimize them accordingly. The accuracy of these approaches is fundamentally constrained by the expressivity of the chosen variational ansatz \cite{PhysRevLett.128.090501, PhysRevLett.127.230501, PhysRevLett.130.236401}.
The third paradigm uses PINNs, which embed the governing physical laws in the loss function \cite{raissi2019physics, karniadakis2021physics}, thereby improving data efficiency for accurate solutions \cite{cuomo2022scientific}. 
While successful in various domains \cite{scienceaaw4741,CaiShengze2021Physics,Fang2020Deep,Raissi2019Wang}, standard PINN architectures face fundamental limitations when applied to systems of coupled DEs: competing gradient updates from DEs can lead to optimization conflicts, and the high-dimensional non-convex loss landscape makes optimization prone to getting trapped in local optima, hindering their accuracy and general applicability \cite{zhang2026physics,yang2025s,cao2025wbpinn}.

To tackle this optimization conflict challenge, we propose a FPINN, a general-purpose computational framework specifically designed for the stable and efficient solution of coupled system of DEs. FPINN adopts a “shared-branch” architecture: the shared component extracts common features across all variables, while the independent branches learn the unique dynamics governed by each individual equation. This design achieves gradient isolation at the computational graph level, effectively mitigating multi-objective optimization conflicts among DEs in coupled problems. 

In this Letter, we demonstrate the FPINN framework through an illustrative application to non-Markovian open quantum dynamics. While PINNs have been adapted for Markovian cases, Ullah et al. \cite{ullah2024physics} enforced trace preservation in the loss function to ensure consistency in dissipative dynamics, and Norambuena et al. \cite{PhysRevLett.132.010801} introduced target and control terms for unified simulation and optimal control, extending them to non-Markovian regimes remains challenging. In this setting, the coupled DEs derived from quantum state diffusion theory \cite{wang2021quantum, diosi1997non, PhysRevLett.83.4909} not only give rise to multi-objective conflicts but also trigger evolutionary stagnation as the network becomes trapped in local optima. FPINN inherently resolves the multi-objective conflicts through its gradient-isolating architecture. To address stagnation, we introduce an evolution regularization term that guides optimization toward dynamic, physically meaningful states. Applications to the spin-boson and XXZ models demonstrate that FPINN successfully overcomes stagnation and accurately reproduces hallmark non-Markovian signatures such as coherence revival and information backflow. The design principles underlying FPINN, together with its demonstrated success in this demanding setting, underscore the framework's intrinsic generality and suggest broad applicability to scientific and engineering problems governed by systems of coupled equations.

\textit{Model---}
For the description of the open quantum system dynamics, an exact non-Markovian master equation for the reduced density matrix has been obtained \cite{wang2021quantum}
\begin{equation}
	\begin{aligned}
		\frac{\partial \rho}{\partial t}
		&=-\:i[H_{s},\rho]+[L,\rho\overline{O}^{\dagger}]-[L^{\dagger},\overline{O}\rho]\\
		&+[L^{\dagger},\rho\overline{Q}^{\dagger}]-[L,\overline{Q}\rho] =\mathcal{F}_{\rho}(\rho,\overline{O},\overline{Q}),
	\end{aligned}
	\label{equ:1}
\end{equation}
where $H_s$ is the system Hamiltonian and $\rho$ is the system density matrix. The auxiliary operators \(\overline{O}(t)=\int_{0}^{t}ds\:\alpha(t-s)O(t,s,z^*_t)\) and \(\overline{Q}(t)=\int_{0}^{t}ds\:\beta(t-s)Q(t,s,z^*_t)\) encapsulate the accumulated environmental memory effects \cite{wang2021quantum}. At time \(t=t_0=0\), the system and bath are uncorrelated, giving \(\overline{O}(t_0)=\overline{Q}(t_0)=0\). Here $\alpha(t-s)$ and $ \beta(t-s) $ denote the correlation functions, see Supplemental Material (SM) for details. The master equation is closed by the dynamical equations for the operators $\overline{O},\overline{Q}$ \cite{wang2021quantum}
\begin{equation}
	\begin{aligned}
		\frac{\partial\overline{O}}{\partial t}
		&=\left(\frac{\Gamma T\gamma}{2}-\frac{i\Gamma\gamma^{2}}{2}\right)L-\gamma\overline{O}+[-iH_s \\
		&-(L^\dagger\overline{O}+L\overline{Q}),\overline{O}] =\mathcal{F}_o(\overline{O},\overline{Q}),
	\end{aligned}
	\label{equ:2}
\end{equation}
\begin{equation}
	\begin{aligned}
		\frac{\partial\overline{Q}}{\partial t}
		&=\frac{\Gamma T\gamma}{2}L^{\dagger}-\gamma\overline{Q}+[-iH_s-(L^\dagger\overline{O}+L\overline{Q}),\overline{Q}]\\
		& =\mathcal{F}_q(\overline{O},\overline{Q}),
	\end{aligned}
	\label{equ:3}
\end{equation}
which propagate $ \overline{O}(t) $ and $ \overline{Q}(t) $ self-consistently. Eqs.~(\ref{equ:2})-(\ref{equ:3}) constitute the coupled DEs system to be solved by the FPINN.
The parameters $ \Gamma $, $ \gamma $ and $ T $ represent the system–bath coupling strength, the bath character frequency, and the temperature, respectively. $ 1/\gamma$ sets the memory time of the environment: $ \gamma \to 0 $ corresponds to colored noise and pronounced memory effects, while $ \gamma\to\infty $ recovers the Markovian regime. 

\begin{figure}[t]
	\centering{\includegraphics[width=\columnwidth]{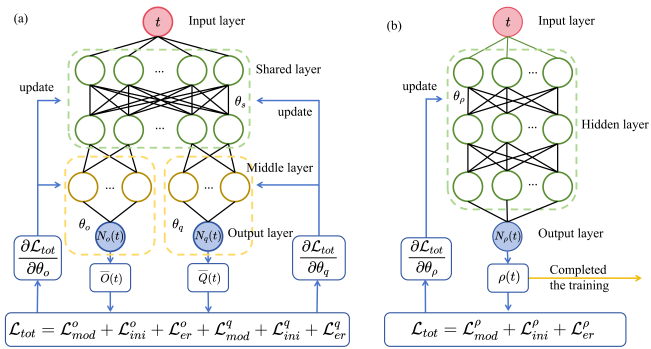}}
	\caption{(a) FPINNs architecture for simulating the dynamics of the auxiliary operators \(\overline{O}(t)\) and \(\overline{Q}(t)\). The red circle represents the input layer with \( t \) as the input. The green dashed part represents the shared layer of FPINN, which consists of three fully connected layers used to learn the common features of different operators. The shared layer is connected to two branched layers (the yellow dashed part), each composed of a middle layer and an output layer, used to learn and output the unique dynamical feature vectors \( N_o(t) \) and \( N_q(t) \) for different operators. (b) PINNs architecture for simulating the dynamics of density matrix \( \rho \). The output layers of PINN and FPINN are the same, with the hidden layer consisting of three fully connected layers, which output the eigenvector \( N_{\rho}(t) \) of the density matrix from the output layer. } 
	\label{fig:1}
\end{figure}

\textit{Architecture---}
The operators \( \overline{O}(t) \) and \( \overline{Q}(t) \) evolve according to the coupled DEs (Eqs.~(\ref{equ:2})-(\ref{equ:2})). While each operator retains its distinct dynamical character, they also share substantial common features. To exploit this structure, we propose a FPINN architecture comprising a shared base network followed by multiple forked branches. The architecture, illustrated in Fig.~\ref{fig:1}(a), begins with an input layer that accepts the time coordinate $ t=(t_0,t_1,\ldots,t_f) $. This is followed by a shared network block $ \theta_s $, constructed from three fully connected layers, which learns the collective physical patterns encoded in the coupled DEs through shared weight parameters. The network then bifurcates into two parallel branches, $ \theta_o $ and $ \theta_q $. Each branch contains a dedicated fully connected hidden layer to extract operator-specific features, and an output layer that produces the corresponding feature vector \( N_{o}(t) \) or \( N_{q}(t) \). The reconstructed operators \( \overline{O}(t) \) and \( \overline{Q}(t) \) are subsequently pooled into a shared computational module to evaluate the total loss $ \mathcal{L}_{tot} $. 

The FPINN training process utilizes PyTorch’s automatic differentiation (AD) framework \cite{paszke2017automatic} to precisely orchestrate the gradient flow. In the forward pass, the input time variable \( t \) first passes through the shared block to capture common features, then propagates separately through the two independent branches to learn operator-specific dynamics, ultimately yielding the operators \( \overline{O}(t) \) or \( \overline{Q}(t) \).
During backpropagation, the gradient paths are explicitly separated: each branch (including its dedicated hidden and output layers) receives gradients only from its corresponding loss term ($\mathcal{L}_{o} $ or $ \mathcal{L}_{q} $), ensuring complete gradient isolation between the two operators. The shared layer, serving as the foundation for both branches, accumulates gradients from both \( \mathcal{L}_{o} \) and \( \mathcal{L}_{q} \), its parameter update is thus the sum of the contributions from the two loss functions. This deliberately structured gradient flow not only preserves the distinct dynamical features of each auxiliary operator, but also through gradient fusion in the shared layer, promotes the extraction of their underlying collective physical patterns.

Fig.~\ref{fig:1}(b) illustrates the workflow for simulating the system density matrix \( \rho \). Here, the network likewise accepts the time variable \( t \) as input. This input is processed through hidden layers to produce a feature vector \( N_\rho(t) \) at the output layer.
Once the networks modeling \( \overline{O} \) and \( \overline{Q} \) have converged, their predictions are supplied as prior knowledge for evaluating the model loss term \( \mathcal{L}_{mod}^\rho \). Upon completion of training, the FPINN provides the full simulated trajectory of the non-Markovian system dynamics.

\textit{Loss function---}
In the training of PINNs, the loss function encodes the physical constraints of the system. The total loss is defined as \cite{raissi2019physics} 
\begin{equation}
	\mathcal{L}_{tot}=\mathcal{L}_{mod}+\mathcal{L}_{ini}+\mathcal{L}_{er},
	\label{equ:4}
\end{equation}
where $\mathcal{L}_{mod}$ enforces the governing dynamical equations, ensuring that the learned evolution obeys the underlying physical laws. $\mathcal{L}_{ini}$ imposes the initial conditions, and $ \mathcal{L}_{er} $ is an evolution regularization term designed to help the network escape from evolutionary stagnation. We employ the FPINN architecture to simulate the dynamics of the operators \(\overline{O}\) and \(\overline{Q}\), while the density matrix \(\rho\) is modeled using a standard PINN. The evolution of the operators governed by the coupled DEs.~(\ref{equ:2}) and (\ref{equ:3}) leads to the following dynamics loss
\begin{equation}
	\mathcal{L}_{mod}^A = \frac{1}{f}\sum_{i=0}^f \bigl\| \frac{\mathrm{d}A(t_i)}{\mathrm{d}t}-\mathcal{F}_A(\overline{O},\overline{Q})\bigr\|^2, 
	\label{equ:5}
\end{equation}	
where \( A \) denotes either the operators \( \overline{O} \), \( \overline{Q} \) or the density matrix \( \rho \). When $A=\rho$, $\mathcal{F}_A(\overline{O},\overline{Q}) $ should be $\mathcal{F}_A(\rho, \overline{O},\overline{Q})$. $\left\|\cdot\right\|$ represents a Euclidean distance. Soft constraints $\mathcal{L}_{ini}^A = \left\|A(t_0) \right\|^2$ (\( A=\overline{O}, \overline{Q} \)) ensures that the simulated $A(t_0)$ is zero \cite{PhysRevLett.132.010801}. 
Once the operators are obtained from the FPINN, they serve as prior knowledge for simulating the dynamics of the density matrix \(\rho(t)\) via Eq.~(\ref{equ:1}). When $A$ is \( \rho \), Eq.~(\ref{equ:5}) becomes the dynamics loss. The initial condition is enforced by $\mathcal{L}_{ini}^{\rho}=\bigl\|\rho(t_0)-\rho_0\bigr\|^2 $, where $\rho_0$ is the desired initial state. 

A common issue encountered when simulating non-Markovian open quantum systems with PINNs is the premature convergence to a static (thermal-equilibrium-like) solution, where the simulated quantities exhibit negligible time dependence. While such solutions can yield small equation residuals, they fail entirely to capture the characteristic time-dependent features of non-Markovian dynamics, such as oscillations and information backflow. To prevent this collapse, we introduce an evolution regularization term \( \mathcal{L}_{er} \) that applies a soft penalty to trajectories with insufficient temporal variation, thereby steering the network toward nontrivial dynamical solutions. The regularization is defined as $	\mathcal{L}_{er}=\lambda_{er}\exp(-\mathrm{TV}[A]/\tau)$. The total variation measure
\begin{equation}
	\mathrm{TV}[A]=\frac{1}{N_t-1}\sum_{i=1}^{N_t-1}\|A(t_i)-A(t_{i-1})\|_1
	\label{equ:6}
\end{equation}
quantifies the average temporal variation of \( A \) over the simulation interval. \( \| \cdot \|_1 \) is the $ L^1 $ norm and \( N_t \) is the number of discrete time points (the normalization makes the measure independent of grid density). The hyperparameter \( \lambda_{er} \) controls the overall penalty strength, while \( \tau \) sets the scale of variation below which the regularization becomes active. The exponential gating \( \exp(-\mathrm{TV}[A]/\tau) \) yields a smooth, always‑nonzero gradient. When $ \mathrm{TV}[A]\ll\tau $, the penalty approaches its maximum value \( \lambda_{er} \), providing a strong driving force to break
evolutionary stagnation. Once the trajectory exhibits sufficient variation $ (\mathrm{TV}[A]\gtrsim\tau) $, the penalty decays rapidly toward zero, automatically ceasing its influence. Consequently, the regularization term supplies a constant ``push” in early training if the sequence is nearly static, yet naturally vanishes in later stages once physically meaningful dynamics emerge. Throughout the optimization, the gradient remains continuous and compatible with standard back propagation based optimizers.

\textit{Results and Discussions---} 
\begin{figure*}[htbp]
	\centering
	\includegraphics[width=1\textwidth]{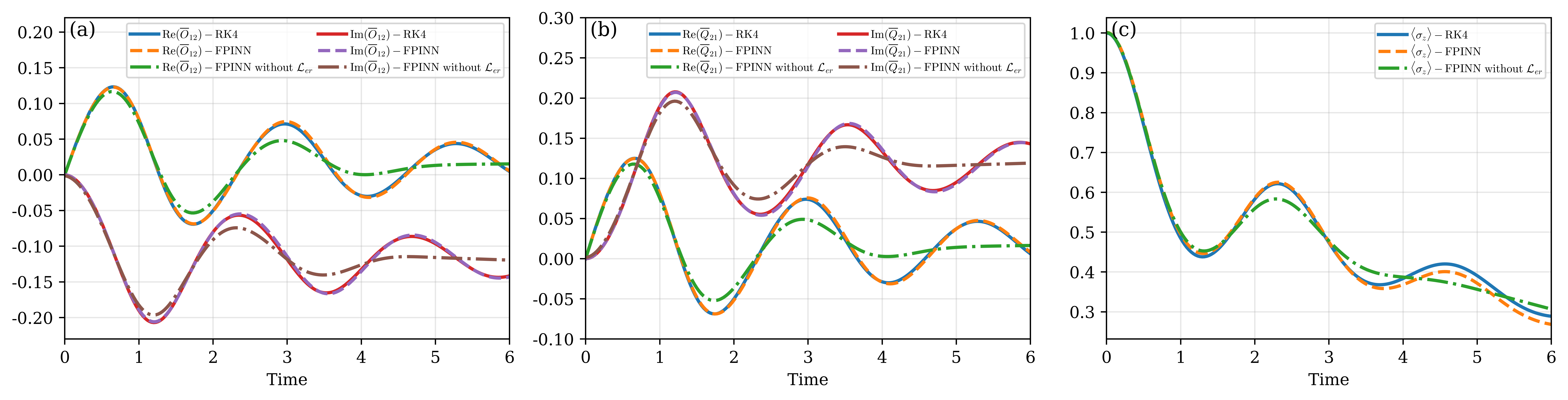}
	\caption{The evolution of the spin-boson model is simulated by RK4 and FPINN. The sampling time points \( t_f = 201 \), the maximum evolution time \( T_{\text{tot}} = 6 \), and the environmental parameters are \( \Gamma = 0.1 \), \( \gamma = 0.3 \), \( T = 20 \). (a), (b), and (c) show the results for \( \overline{O}_{12} \), \( \overline{Q}_{21} \), and \( \langle \sigma_z \rangle \), respectively.} 
	\label{fig:2}
\end{figure*}
To demonstrate the effectiveness of the FPINN framework, we examine the dynamics of a two level spin‑boson model \cite{PhysRevA.86.012115,PhysRevA.104.012213} in a non‑Markovian heat bath. $H_s = \sigma_z $,  $L=\sigma_x $, and the initial state is $|0\rangle$. Fig.~\ref{fig:2} presents the simulated time evolution of the operators \(\overline{O}(t)\), \(\overline{Q}(t)\) and the density matrix \(\rho(t)\). For clarity, we only plot selected matrix elements of the operators (See SM for the full results). The information backflow due to the memory effects of the environment manifests as a gradually damped oscillation in the $\langle\sigma_z\rangle$ value, indicating our FPINN framework has faithflly captured this hallmark of non-Markovian dynamics \cite{PhysRevA.104.012213}. The simulated evolution shows excellent agreement with the reference solution obtained via a fourth-order Runge–Kutta (RK4) method. Quantitatively, the average Frobenius-norm error over all sampled time points is only about $ 0.3\%$, further corroborating the effectiveness of FPINN in simulating non-Markovian dynamics.

Next we compare the FPINN simulations with and without the evolution regularization loss term \( \mathcal{L}_{er} \). As shown in Fig.~\ref{fig:2}, the model trained without \( \mathcal{L}_{er} \) tends to produce trajectories that prematurely stagnate in the latter time, deviating substantially from the reference RK4 results. This behavior stems from the network’s tendency to settle into ``lazy” local optima that yield nearly static solutions with small equation residuals but miss the essential features of non-Markovian dynamics. The inclusion of \( \mathcal{L}_{er} \) penalizes such insufficient temporal variation and effectively pushes the optimization away from these trivial equilibria, leading to solutions that successfully reproduce characteristic non‑Markovian signatures. We further note that in more Markovian regimes, the simulations obtained without \( \mathcal{L}_{er} \) also become increasingly accurate, as detailed in the SM.

\begin{figure}[htbp]
	\centering{\includegraphics[width=\columnwidth]{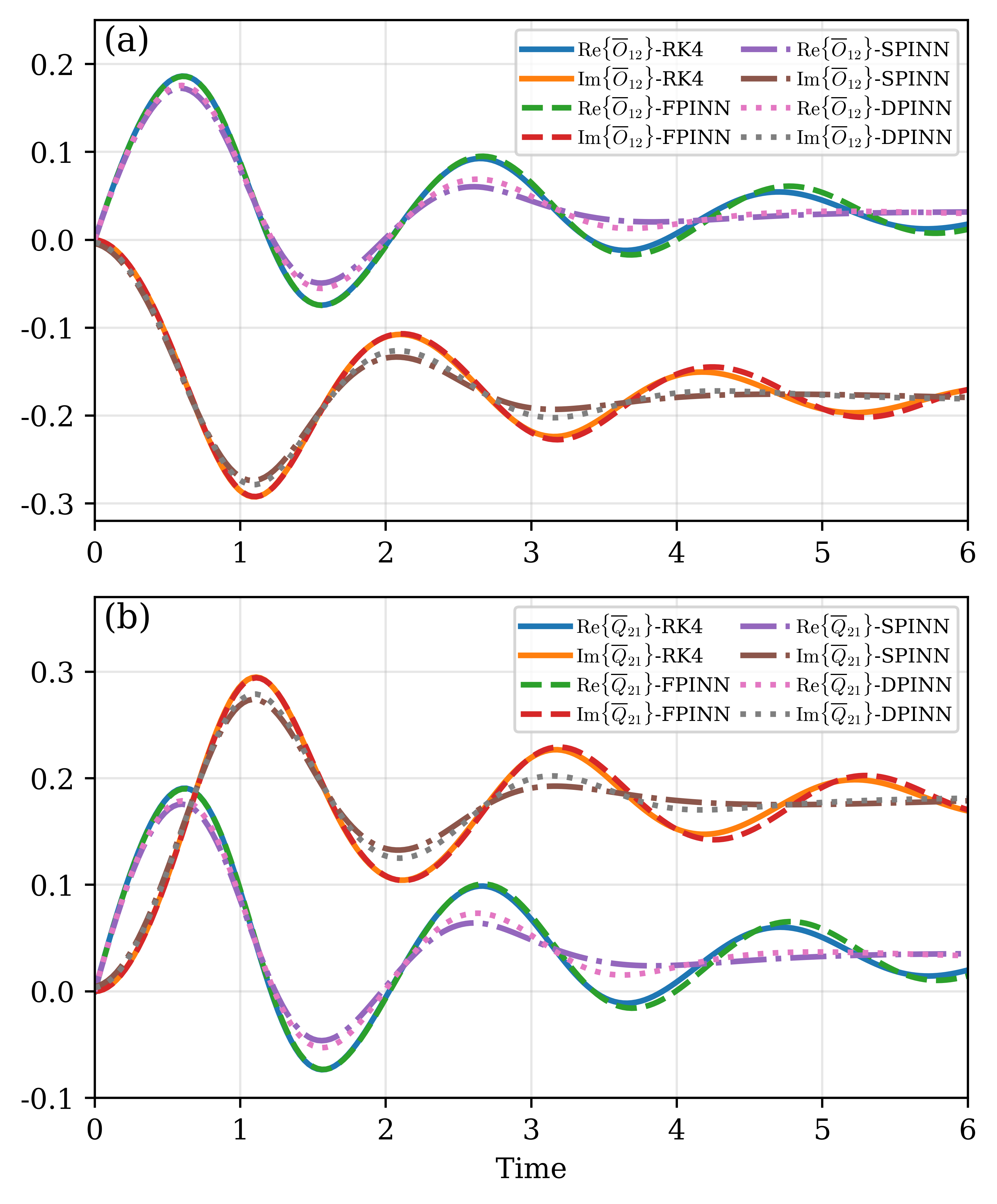}}
	\caption{(a) and (b) show the evolution of the auxiliary operators \( \overline{O} \) and \( \overline{Q} \) for the spin-boson model simulated by RK4, FPINN, UPINN, and SPINN, respectively. \( \Gamma = 0.1 \), \( \gamma = 0.5 \), \( T = 20 \).} 
	\label{fig:3}
\end{figure}

We also evaluated FPINN against two alternative PINN architectures designed for the same task. The Unified PINN (UPINN) employs a single network that outputs all components of \( \overline{O}(t) \) and \( \overline{Q}(t) \) simultaneously, hidden layers are fully shared, with separation only at the output layer. In contrast, the Separated PINN (SPINN) consists of two independent, identically structured networks, each dedicated to simulating either \( \overline{O}(t) \) and \( \overline{Q}(t) \). A comparative analysis of the three architectures is presented in Fig.~\ref{fig:3}. The FPINN predictions align significantly better with the RK4 reference solutions than those of UPINN or SPINN. The UPINN architecture suffers from  multi-objective optimization conflicts \cite{yang2025s}, as the competing objectives for \( \overline{O} \) and \( \overline{Q} \) contend for limited network capacity, often leading to compromised accuracy. The SPINN approach, while avoiding such conflict, disregards the intrinsic coupling between the two operators encoded in the physical equations, which can result in unphysical or uncoordinated feature learning.
FPINN addresses both limitations: its shared layers capture the common dynamical patterns arising from the operator coupling, while the dedicated branches adapt to operator‑specific variations. This hybrid design makes FPINN particularly well‑suited for solving the coupled DEs that describe non‑Markovian auxiliary operators. Moreover, the architecture can be naturally extended to other scenarios requiring the joint solution of coupled DEs.

To further validate the flexibility and accuracy of the FPINN architecture, we apply it to a dissipative two qubit Heisenberg XXZ model, a paradigmatic many body system widely used to study spin interactions \cite{yo2025generation,PhysRevA.72.034302,CAI20104415}. $ H = J(\sigma_x^1\sigma_x^2 + \sigma_y^1\sigma_y^2) + \Delta(\sigma_z^1\sigma_z^2)$. $J$ and $\Delta$ denote the in-plane and out-of-plane coupling strengths, respectively, with $J = 2$ and $\Delta = 0.5$, corresponding to antiferromagnetic coupling ($J>0$) and easy-plane anisotropy ($|\Delta|/J<1$) dominated by $xy$-plane interactions. Dissipation is introduced through a collective Lindblad operator $L =\sum_{i=1}^{2}\sigma_i^-$, where $\sigma_{i}^{-}$ denotes the lowering operator. The operators \( \overline{O} \) and \( \overline{Q} \) encode the environmental memory and are independent of the system’s initial state. Hence, once simulated for a given environment, they can be reused to compute the dynamics for different initial conditions. We examine two representative initial states: \( |00\rangle \) and the Bell state \( (|00\rangle + |11\rangle) /\sqrt{2}\). For this system, we calculate the time evolution of coherence and concurrence. The coherence is quantified by the \( l_1 \) norm \cite{PhysRevLett.113.140401, RevModPhys.89.041003}, \( C(t) = \sum_{i \neq j} |\rho(t)_{i,j}| \), where $ \rho_{ij}(t)=\langle i|\rho(t)|j\rangle $ and \( i \), \( j \) label the two qubit energy eigenstates. Entanglement is measured by the concurrence \cite{PhysRevLett.80.2245}, \(\mathcal{C}(\rho) = \max \left\{0, \sqrt{\lambda_{1}} - \sqrt{\lambda_{2}} - \sqrt{\lambda_{3}} - \sqrt{\lambda_{4}} \right\}\) with \(\lambda_{1} \geq \lambda_{2} \geq \lambda_{3} \geq \lambda_{4}\) the eigenvalues of \(R = \rho (\sigma_{y} \otimes \sigma_{y}) \rho^{*} (\sigma_{y} \otimes \sigma_{y})\). 

\begin{figure}[htbp]
	\centering{\includegraphics[width=\columnwidth]{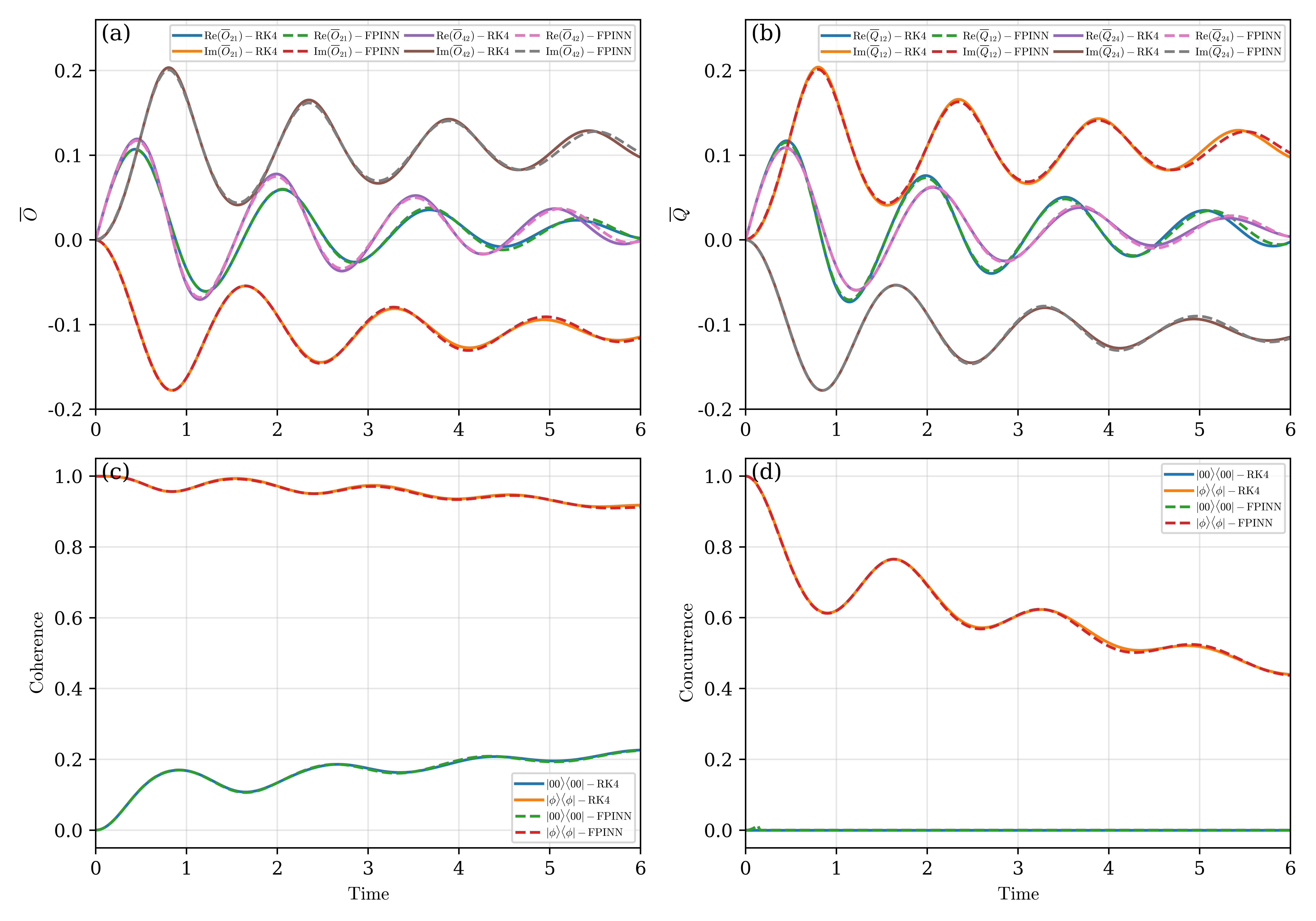}}
	\caption{The evolution of the XXZ model is simulated by RK4 and FPINN. The sampling time points \( t_f = 401 \), the maximum evolution time \( T_{\text{tot}} = 6 \), and the environmental parameters are \( \Gamma = 0.1 \), \( \gamma = 0.4 \), \( T = 20 \). (a) and (b) show the evolution of the auxiliary operators \( O \) and \( Q \), respectively, while (c) and (d) show the evolution of quantum coherence and quantum entanglement for different initial states.} 
	\label{fig:4}
\end{figure}

Fig.~\ref{fig:4}(a) and (b) display the FPINN‑simulated trajectories of \( \overline{O} \) and \( \overline{Q} \), while Fig.~\ref{fig:4}(c) and (d) show the resulting coherence and entanglement dynamics for the two initial states using the same operators. For the state \( |00\rangle \), under the combined influence of the Hamiltonian and the non‑Markovian environment, oscillatory coherence emerges from zero while entanglement remains absent. 
In contrast, for the Bell state, both quantites decay over time with initial unity under both quantities decay over time under environmental dissipation, exhibiting clear non‑Markovian backflow revivals. The FPINN simulations accurately capture these many‑body dynamical features, further demonstrating the method’s capability to handle non-Markovian many-body open quantum system dynamics.

\textit{Conclusions---} 
We propose an effective FPINN architecture to address the multi-objective optimization conflicts in coupled systems of DEs. As an illustrative example, we simulate non-Markovian open quantum dynamics with our framework, requiring solution of coupled operators. The network, through the collaborative design of shared and branched layers, effectively isolates gradient competition of the branched layers at the computational graph level while learning the common features in the coupling relationships between operators, thereby enhancing training stability. 
The proposed evolution regularization loss function further prevents the network from falling into local optima. Simulations on the spin-boson and XXZ models demonstrate that FPINN significantly outperforms traditional UPINN and SPINN architectures, accurately simulating coherent revival and information backflow in non-Markovian dynamics and achieving highly consistent results with those obtained by RK4 method. FPINN provides an efficient and general deep learning framework for simulating coupled systems of DEs, and can be extended to address other types of coupled equations, such as the Hartree–Fock \cite{Comparison2000Petr} and Kohn–Sham \cite{Abdulrahman2026Kohn} equations in wave-function-based quantum chemistry and density functional theory, respectively.

\section{Acknowledgments}
This paper is supported by the Natural Science Foundation of Shandong Province (Grant No.ZR2024MA046). 

\FloatBarrier
\bibliography{References_library.bib}

@article{cuomo2022scientific,
  title={Scientific machine learning through physics--informed neural networks: Where we are and what’s next},
  author={Cuomo, Salvatore and Di Cola, Vincenzo Schiano and Giampaolo, Fabio and Rozza, Gianluigi and Raissi, Maziar and Piccialli, Francesco},
  journal={Journal of Scientific Computing},
  volume={92},
  number={3},
  pages={88},
  year={2022},
  publisher={Springer},
  url = {https://doi.org/10.1007/s10915-022-01939-z},
  doi = {10.1007/s10915-022-01939-z}
}

@article{cai2021physics,
  title={Physics-informed neural networks (PINNs) for fluid mechanics: A review},
  author={Cai, Shengze and Mao, Zhiping and Wang, Zhicheng and Yin, Minglang and Karniadakis, George Em},
  journal={Acta Mechanica Sinica},
  volume={37},
  number={12},
  pages={1727--1738},
  year={2021},
  publisher={Springer},
  url = {https://doi.org/10.1007/s10409-021-01148-1},
  doi = {10.1007/s10409-021-01148-1}
}

@article{MAO2020112789,
title = {Physics-informed neural networks for high-speed flows},
journal = {Computer Methods in Applied Mechanics and Engineering},
volume = {360},
pages = {112789},
year = {2020},
issn = {0045-7825},
doi = {https://doi.org/10.1016/j.cma.2019.112789},
url = {https://www.sciencedirect.com/science/article/pii/S0045782519306814},
author = {Zhiping Mao and Ameya D. Jagtap and George Em Karniadakis}
}

@article{PhysRevLett.132.010801,
  title = {Physics-Informed Neural Networks for Quantum Control},
  author = {Norambuena, Ariel and Mattheakis, Marios and Gonz\'alez, Francisco J. and Coto, Ra\'ul},
  journal = {Phys. Rev. Lett.},
  volume = {132},
  issue = {1},
  pages = {010801},
  numpages = {7},
  year = {2024},
  month = {Jan},
  publisher = {American Physical Society},
  doi = {10.1103/PhysRevLett.132.010801},
  url = {https://link.aps.org/doi/10.1103/PhysRevLett.132.010801}
}

@article{ullah2024physics,
  title={Physics-informed neural networks and beyond: enforcing physical constraints in quantum dissipative dynamics},
  author={Ullah, Arif and Huang, Yu and Yang, Ming and Dral, Pavlo O},
  journal={Digital Discovery},
  volume={3},
  number={10},
  pages={2052--2060},
  year={2024},
  publisher={Royal Society of Chemistry},
  doi = {10.1039/D4DD00153B},
  url = {https://doi.org/10.1039/D4DD00153B}
}

@article{karniadakis2021physics,
  title={Physics-informed machine learning},
  author={Karniadakis, George Em and Kevrekidis, Ioannis G and Lu, Lu and Perdikaris, Paris and Wang, Sifan and Yang, Liu},
  journal={Nature Reviews Physics},
  volume={3},
  number={6},
  pages={422--440},
  year={2021},
  publisher={Nature Publishing Group UK London},
url = {https://doi.org/10.1038/s42254-021-00314-5},
doi={10.1038/s42254-021-00314-5}
}

@article{PhysRevLett.83.4909,
  title = {Quantum Trajectories for Brownian Motion},
  author = {Strunz, Walter T. and Di\'osi, Lajos and Gisin, Nicolas and Yu, Ting},
  journal = {Phys. Rev. Lett.},
  volume = {83},
  issue = {24},
  pages = {4909--4913},
  numpages = {0},
  year = {1999},
  month = {Dec},
  publisher = {American Physical Society},
  doi = {10.1103/PhysRevLett.83.4909},
  url = {https://link.aps.org/doi/10.1103/PhysRevLett.83.4909}
}

@article{WANG201078,
title = {From coherent motion to localization: II. Dynamics of the spin-boson model with sub-Ohmic spectral density at zero temperature},
journal = {Chemical Physics},
volume = {370},
number = {1},
pages = {78-86},
year = {2010},
note = {Dynamics of molecular systems: From quantum to classical},
issn = {0301-0104},
doi = {https://doi.org/10.1016/j.chemphys.2010.02.027},
url = {https://www.sciencedirect.com/science/article/pii/S0301010410001011},
author = {Haobin Wang and Michael Thoss},
}

@article{ritschel2014analytic,
  title={Analytic representations of bath correlation functions for ohmic and superohmic spectral densities using simple poles},
  author={Ritschel, Gerhard and Eisfeld, Alexander},
  journal={J. Chem. Phys.},
  volume={141},
  number={9},
  year={2014},
  doi = {10.1063/1.4893931},
  url = {https://doi.org/10.1063/1.4893931},
  publisher={AIP Publishing}
}

@article{PhysRevLett.127.230501,
  title = {Time-Dependent Variational Principle for Open Quantum Systems with Artificial Neural Networks},
  author = {Reh, Moritz and Schmitt, Markus and G\"arttner, Martin},
  journal = {Phys. Rev. Lett.},
  volume = {127},
  issue = {23},
  pages = {230501},
  numpages = {7},
  year = {2021},
  month = {Dec},
  publisher = {American Physical Society},
  doi = {10.1103/PhysRevLett.127.230501},
  url = {https://link.aps.org/doi/10.1103/PhysRevLett.127.230501}
}

@article{PhysRevLett.128.090501,
  title = {Autoregressive Neural Network for Simulating Open Quantum Systems via a Probabilistic Formulation},
  author = {Luo, Di and Chen, Zhuo and Carrasquilla, Juan and Clark, Bryan K.},
  journal = {Phys. Rev. Lett.},
  volume = {128},
  issue = {9},
  pages = {090501},
  numpages = {7},
  year = {2022},
  month = {Feb},
  publisher = {American Physical Society},
  doi = {10.1103/PhysRevLett.128.090501},
  url = {https://link.aps.org/doi/10.1103/PhysRevLett.128.090501}
}

@article{PhysRevA.72.034302,
  title = {Thermal entanglement in a two-qubit Heisenberg $XXZ$ spin chain under an inhomogeneous magnetic field},
  author = {Zhang, Guo-Feng and Li, Shu-Shen},
  journal = {Phys. Rev. A},
  volume = {72},
  issue = {3},
  pages = {034302},
  numpages = {4},
  year = {2005},
  month = {Sep},
  publisher = {American Physical Society},
  doi = {10.1103/PhysRevA.72.034302},
  url = {https://link.aps.org/doi/10.1103/PhysRevA.72.034302}
}

@article{CAI20104415,
author = {Jiang-Tao Cai and Ahmad Abliz and Guo-Feng Zhang and Yan-Kui Bai},
title = {Effects of Dzyaloshinskii–Moriya interaction on thermal entanglement and teleportation via a two-qubit Heisenberg XXZ spin chain under external magnetic field},
journal = {Optics Communications},
volume = {283},
number = {21},
pages = {4415-4421},
year = {2010},
issn = {0030-4018},
doi = {https://doi.org/10.1016/j.optcom.2010.06.075},
url = {https://www.sciencedirect.com/science/article/pii/S0030401810006875},
}

@article{yo2025generation,
  title={Generation of Maximally Entanglement States by Quantum Particle Swarm Optimization Under the Decoherence Channel in the Two-Qubit Heisenberg XXZ Model with DM and KSEA Interaction},
  author={Yo, Cholmyong and Jon, Song and Kang, Unil and Jyongyon, Kim and Jon, Hyok},
  journal={International Journal of Theoretical Physics},
  volume={64},
  number={5},
  pages={149},
  year={2025},
  publisher={Springer},
doi = {10.1007/s10773-025-05994-8},
url = {https://doi.org/10.1007/s10773-025-05994-8},
}

@article{PhysRevLett.80.2245,
  title = {Entanglement of Formation of an Arbitrary State of Two Qubits},
  author = {Wootters, William K.},
  journal = {Phys. Rev. Lett.},
  volume = {80},
  issue = {10},
  pages = {2245--2248},
  numpages = {0},
  year = {1998},
  month = {Mar},
  publisher = {American Physical Society},
  doi = {10.1103/PhysRevLett.80.2245},
  url = {https://link.aps.org/doi/10.1103/PhysRevLett.80.2245}
}

@article{raissi2019physics,
  title={Physics-informed neural networks: A deep learning framework for solving forward and inverse problems involving nonlinear partial differential equations},
  author={Raissi, Maziar and Perdikaris, Paris and Karniadakis, George E},
  journal={Journal of Computational physics},
  volume={378},
  pages={686--707},
  year={2019},
  publisher={Elsevier},
  doi = {10.1016/j.jcp.2018.10.045},
  url = {https://doi.org/10.1016/j.jcp.2018.10.045}
}

@article{diosi1997non,
  title={The non-Markovian stochastic Schr{\"o}dinger equation for open systems},
  author={Di{\'o}si, Lajos and Strunz, Walter T},
  journal={Physics Letters A},
  volume={235},
  number={6},
  pages={569--573},
  year={1997},
  publisher={Elsevier},
  doi = {10.1016/S0375-9601(97)00717-2},
  url = {https://doi.org/10.1016/S0375-9601(97)00717-2}
}

@article{wang2021quantum,
  title={Quantum state transmission through a spin chain in finite-temperature heat baths},
  author={Wang, Zhao-Ming and Ren, Feng-Hua and Luo, Da-Wei and Yan, Zhan-Yuan and Wu, Lian-Ao},
  journal={Journal of Physics A: Mathematical and Theoretical},
  volume={54},
  number={15},
  pages={155303},
  year={2021},
  publisher={IOP Publishing},
  doi = {10.1088/1751-8121/abe751},
  url = {https://doi.org/10.1088/1751-8121/abe751}
}

@article{yang2025s,
  title={S-PINN: Stabilized physics-informed neural networks for alleviating barriers between multi-level co-optimization},
  author={Yang, Tengmao and Qian, Zhihao and Hang, Nianzhi and Liu, Moubin},
  journal={Computer Methods in Applied Mechanics and Engineering},
  volume={447},
  pages={118348},
  year={2025},
  publisher={Elsevier},
  doi = {10.1016/j.cma.2025.118348},
  url = {https://doi.org/10.1016/j.cma.2025.118348}
}

@misc{paszke2017automatic,
  title={Automatic differentiation in pytorch},
  author={Paszke, Adam and Gross, Sam and Chintala, Soumith and Chanan, Gregory and Yang, Edward and DeVito, Zachary and Lin, Zeming and Desmaison, Alban and Antiga, Luca and Lerer, Adam},
  year={2017},
month = {10},
howpublished = {OpenReview},
 url = {https://openreview.net/forum?id=BJJsrmfCZ},
}

@article{RevModPhys.89.041003,
  title = {Colloquium: Quantum coherence as a resource},
  author = {Streltsov, Alexander and Adesso, Gerardo and Plenio, Martin B.},
  journal = {Rev. Mod. Phys.},
  volume = {89},
  issue = {4},
  pages = {041003},
  numpages = {34},
  year = {2017},
  month = {Oct},
  publisher = {American Physical Society},
  doi = {10.1103/RevModPhys.89.041003},
  url = {https://link.aps.org/doi/10.1103/RevModPhys.89.041003}
}

@article{PhysRevLett.113.140401,
  title = {Quantifying Coherence},
  author = {Baumgratz, T. and Cramer, M. and Plenio, M. B.},
  journal = {Phys. Rev. Lett.},
  volume = {113},
  issue = {14},
  pages = {140401},
  numpages = {5},
  year = {2014},
  month = {Sep},
  publisher = {American Physical Society},
  doi = {10.1103/PhysRevLett.113.140401},
  url = {https://link.aps.org/doi/10.1103/PhysRevLett.113.140401}
}

@article{shukla2021parallel,
  title={Parallel physics-informed neural networks via domain decomposition},
  author={Shukla, Khemraj and Jagtap, Ameya D and Karniadakis, George Em},
  journal={Journal of Computational Physics},
  volume={447},
  pages={110683},
  year={2021},
  publisher={Elsevier},
doi = {https://doi.org/10.1016/j.jcp.2021.110683},
url = {https://www.sciencedirect.com/science/article/pii/S0021999121005787},
}

@article{tanimura2020numerically,
  title={Numerically “exact” approach to open quantum dynamics: The hierarchical equations of motion (HEOM)},
  author={Tanimura, Yoshitaka},
  journal={The Journal of chemical physics},
  volume={153},
  number={2},
  year={2020},
  publisher={AIP Publishing},
doi = {10.1063/5.0011599},
url = {https://doi.org/10.1063/5.0011599},
}

@article{bai2024hierarchical,
  title={Hierarchical equations of motion for quantum chemical dynamics: Recent methodology developments and applications},
  author={Bai, Shuming and Zhang, Shuocang and Huang, Chenghong and Shi, Qiang},
  journal={Accounts of Chemical Research},
  volume={57},
  number={21},
  pages={3151--3160},
  year={2024},
  publisher={ACS Publications},
doi={10.1021/acs.accounts.4c00492},
url={https://doi.org/10.1021/acs.accounts.4c00492}
}

@article{cao2025wbpinn,
  title={wbPINN: Weight balanced physics-informed neural networks for multi-objective learning},
  author={Cao, Fujun and Guo, Xiaobin and Dong, Xinzheng and Yuan, Dongfang},
  journal={Applied Soft Computing},
  volume={170},
  pages={112632},
  year={2025},
  publisher={Elsevier},
doi = {https://doi.org/10.1016/j.asoc.2024.112632},
url = {https://www.sciencedirect.com/science/article/pii/S1568494624014066},
}

@article{PhysRevA.86.012115,
  title = {Quantification of memory effects in the spin-boson model},
  author = {Clos, Govinda and Breuer, Heinz-Peter},
  journal = {Phys. Rev. A},
  volume = {86},
  issue = {1},
  pages = {012115},
  numpages = {6},
  year = {2012},
  month = {Jul},
  publisher = {American Physical Society},
  doi = {10.1103/PhysRevA.86.012115},
  url = {https://link.aps.org/doi/10.1103/PhysRevA.86.012115}
}

@article{PhysRevA.104.012213,
  title = {Non-Markovian effects in the spin-boson model at zero temperature},
  author = {Wenderoth, S. and Breuer, H.-P. and Thoss, M.},
  journal = {Phys. Rev. A},
  volume = {104},
  issue = {1},
  pages = {012213},
  numpages = {9},
  year = {2021},
  month = {Jul},
  publisher = {American Physical Society},
  doi = {10.1103/PhysRevA.104.012213},
  url = {https://link.aps.org/doi/10.1103/PhysRevA.104.012213}
}

@article{PhysRevLett.130.236401,
  title = {Transformer Variational Wave Functions for Frustrated Quantum Spin Systems},
  author = {Viteritti, Luciano Loris and Rende, Riccardo and Becca, Federico},
  journal = {Phys. Rev. Lett.},
  volume = {130},
  issue = {23},
  pages = {236401},
  numpages = {6},
  year = {2023},
  month = {Jun},
  publisher = {American Physical Society},
  doi = {10.1103/PhysRevLett.130.236401},
  url = {https://link.aps.org/doi/10.1103/PhysRevLett.130.236401}
}

@article{han2019review,
  title={A review of deep learning models for time series prediction},
  author={Han, Zhongyang and Zhao, Jun and Leung, Henry and Ma, King Fai and Wang, Wei},
  journal={IEEE Sensors Journal},
  volume={21},
  number={6},
  pages={7833--7848},
  year={2019},
  publisher={IEEE},
  doi={10.1109/JSEN.2019.2923982},
  url = {https://ieeexplore.ieee.org/abstract/document/8742529}
}

@article{lin2021simulation,
  title={Simulation of open quantum dynamics with bootstrap-based long short-term memory recurrent neural network},
  author={Lin, Kunni and Peng, Jiawei and Gu, Feng Long and Lan, Zhenggang},
  journal={The Journal of Physical Chemistry Letters},
  volume={12},
  number={41},
  pages={10225--10234},
  year={2021},
  publisher={ACS Publications},
 doi = {10.1021/acs.jpclett.1c02672},
 url = {https://doi.org/10.1021/acs.jpclett.1c02672}
}

@article{wu2021forecasting,
  title={Forecasting nonadiabatic dynamics using hybrid convolutional neural network/long short-term memory network},
  author={Wu, Daxin and Hu, Zhubin and Li, Jiebo and Sun, Xiang},
  journal={The Journal of Chemical Physics},
  volume={155},
  number={22},
  year={2021},
  publisher={AIP Publishing},
doi = {10.1063/5.0073689},
url = {https://doi.org/10.1063/5.0073689}
}

@article{NIEGEMANN2012364,
title = {Efficient low-storage Runge–Kutta schemes with optimized stability regions},
journal = {Journal of Computational Physics},
volume = {231},
number = {2},
pages = {364-372},
year = {2012},
issn = {0021-9991},
doi = {https://doi.org/10.1016/j.jcp.2011.09.003},
url = {https://www.sciencedirect.com/science/article/pii/S0021999111005213},
author = {Jens Niegemann and Richard Diehl and Kurt Busch},
}

@article{LIU2024109055,
title = {Mass and energy conservative high-order diagonally implicit Runge–Kutta schemes for nonlinear Schrödinger equation},
journal = {Applied Mathematics Letters},
volume = {153},
pages = {109055},
year = {2024},
issn = {0893-9659},
doi = {https://doi.org/10.1016/j.aml.2024.109055},
url = {https://www.sciencedirect.com/science/article/pii/S0893965924000752},
author = {Ziyuan Liu and Hong Zhang and Xu Qian and Songhe Song},
}

@article{Comparison2000Petr,
author = {Bouř, Petr},
title = {Comparison of Hartree–Fock and Kohn–Sham determinants as wave functions},
journal = {Journal of Computational Chemistry},
volume = {21},
number = {1},
pages = {8-16},
doi = {https://doi.org/10.1002/(SICI)1096-987X(20000115)21:1<8::AID-JCC2>3.0.CO;2-Q},
url = {https://onlinelibrary.wiley.com/doi/abs/10.1002/%28SICI%291096-987X%2820000115%2921%3A1%3C8%3A%3AAID-JCC2%3E3.0.CO%3B2-Q},
year = {2000}
}

@article{zhang2026physics,
  title  = {Physics-informed neural networks (PINNs) as intelligent computing technique for solving partial differential equations: Limitation and future prospects},
  author = {Zhang, Weiwei and Suo, Wei and Song, Jiahao and Cao, Wenbo},
  journal = {Science China Physics, Mechanics \& Astronomy},
  volume = {69},
  number = {1},
  pages = {214602},
  year = {2026},
  publisher = {Science China Press},
  doi = {10.1007/s11433-024-2665-5},
  url = {https://doi.org/10.1007/s11433-024-2665-5}
}

@article{courant1928differenzengleichungen,
  title = {Über die partiellen Differenzengleichungen der mathematischen Physik},
  author = {Courant, R. and Friedrichs, K. and Lewy, H.},
  journal = {Mathematische Annalen},
  volume = {100},
  number = {1},
  pages = {32--74},
  year = {1928},
  publisher = {Springer-Verlag},
  doi = {10.1007/BF01448839},
  url = {https://doi.org/10.1007/BF01448839}
}

@article{STYNES199583,
title = {Finite volume methods for convection-diffusion problems},
journal = {Journal of Computational and Applied Mathematics},
volume = {63},
number = {1},
pages = {83-90},
year = {1995},
note = {Proceedings of the International Symposium on Mathematical Modelling and Computational Methods Modelling 94},
issn = {0377-0427},
doi = {https://doi.org/10.1016/0377-0427(95)00056-9},
url = {https://www.sciencedirect.com/science/article/pii/0377042795000569},
author = {Martin Stynes}
}

@article{DIVO2005136,
title = {A meshless method for conjugate heat transfer problems},
journal = {Engineering Analysis with Boundary Elements},
volume = {29},
number = {2},
pages = {136-149},
year = {2005},
issn = {0955-7997},
doi = {https://doi.org/10.1016/j.enganabound.2004.10.001},
url = {https://www.sciencedirect.com/science/article/pii/S0955799704001584},
author = {Eduardo Divo and Alain J. Kassab},
}

@article{Zerroukat2000Explicit,
author = {Zerroukat, M. and Djidjeli, K. and Charafi, A.},
title = {Explicit and implicit meshless methods for linear advection–diffusion-type partial differential equations},
journal = {International Journal for Numerical Methods in Engineering},
volume = {48},
number = {1},
pages = {19-35},
doi = {https://doi.org/10.1002/(SICI)1097-0207(20000510)48:1<19::AID-NME862>3.0.CO;2-3},
url = {https://onlinelibrary.wiley.com/doi/abs/10.1002/%28SICI%291097-0207%2820000510%2948%3A1%3C19%3A%3AAID-NME862%3E3.0.CO%3B2-3},
year = {2000}
}

@article{DOOSTAN20094332,
title = {A least-squares approximation of partial differential equations with high-dimensional random inputs},
journal = {Journal of Computational Physics},
volume = {228},
number = {12},
pages = {4332-4345},
year = {2009},
issn = {0021-9991},
doi = {https://doi.org/10.1016/j.jcp.2009.03.006},
url = {https://www.sciencedirect.com/science/article/pii/S0021999109001144},
author = {Alireza Doostan and Gianluca Iaccarino},
}

@article{scienceaaw4741,
author = {Maziar Raissi  and Alireza Yazdani  and George Em Karniadakis },
title = {Hidden fluid mechanics: Learning velocity and pressure fields from flow visualizations},
journal = {Science},
volume = {367},
number = {6481},
pages = {1026-1030},
year = {2020},
doi = {10.1126/science.aaw4741},
URL = {https://www.science.org/doi/abs/10.1126/science.aaw4741},
}

@article{CaiShengze2021Physics,
    author = {Cai, Shengze and Wang, Zhicheng and Wang, Sifan and Perdikaris, Paris and Karniadakis, George Em},
    title = {Physics-Informed Neural Networks for Heat Transfer Problems},
    journal = {Journal of Heat Transfer},
    volume = {143},
    number = {6},
    pages = {060801},
    year = {2021},
    month = {04},
    issn = {0022-1481},
    doi = {10.1115/1.4050542},
    url = {https://doi.org/10.1115/1.4050542},
}

@article{Fang2020Deep,
  author={Fang, Zhiwei and Zhan, Justin},
  journal={IEEE Access}, 
  title={Deep Physical Informed Neural Networks for Metamaterial Design}, 
  year={2020},
  volume={8},
  number={},
  pages={24506-24513},
  doi={10.1109/ACCESS.2019.2963375}
}

@article{Raissi2019Wang, 
title={Deep learning of vortex-induced vibrations}, 
volume={861}, 
doi={10.1017/jfm.2018.872}, 
journal={Journal of Fluid Mechanics},
 author={Raissi, Maziar and Wang, Zhicheng and Triantafyllou, Michael S. and Karniadakis, George Em},
 year={2019},
 pages={119–137}
}

@misc{Abdulrahman2026Kohn,
  title={Kohn-Sham density encoding rescues coupled cluster theory for strongly correlated molecules},
  author={Abdulrahman Y. Zamani and Barbaro Zulueta and Andrew M. Ricciuti and John A. Keith and Kevin Carter-Fenk},
  journal={arXiv preprint arXiv:2602.06149},
  year={2026},
  doi = {arXiv:2602.06149},
  url = {https://doi.org/10.48550/arXiv.2602.06149}
}

@misc{JunDong2026Optimal,
  title={Optimal Control Design Guided by Adam Algorithm and LSTM-Predicted Open Quantum System Dynamics},
  author={JunDong Zhong, ZhaoMing Wang},
  journal={arXiv preprint arXiv:2602.04480},
  year={2026},
  doi = {arXiv:2602.04480},
  url = {https://doi.org/10.48550/arXiv.2602.04480}
}

@article{PhysRevA.58.1699,
  title = {Non-Markovian quantum state diffusion},
  author = {Di\'osi, L. and Gisin, N. and Strunz, W. T.},
  journal = {Phys. Rev. A},
  volume = {58},
  issue = {3},
  pages = {1699--1712},
  numpages = {0},
  year = {1998},
  month = {Sep},
  publisher = {American Physical Society},
  doi = {10.1103/PhysRevA.58.1699},
  url = {https://link.aps.org/doi/10.1103/PhysRevA.58.1699}
}

@misc{loshchilov2017decoupled,
  title={Decoupled weight decay regularization},
  author={Loshchilov, Ilya and Hutter, Frank},
  journal={arXiv preprint arXiv:1711.05101},
  year={2017},
 doi = {arXiv.1711.05101},
 url = {https://doi.org/10.48550/arXiv.1711.05101}
}

@misc{loshchilov2016sgdr,
  title={Sgdr: Stochastic gradient descent with warm restarts},
  author={Loshchilov, Ilya and Hutter, Frank},
  journal={arXiv preprint arXiv:1608.03983},
  year={2016},
 url ={https://doi.org/10.48550/arXiv.1608.03983},
 doi = {arXiv.1608.03983}
}

\clearpage
\section*{Supplemental Material: ``Forked Physics Informed Neural Networks for Coupled Systems of Differential equations''}
\section{The derivation of the non-Markovian master equation}
The total Hamiltonian of the open quantum system is $ H_{tot} = H_s + H_b + H_{int} $, where $ H_s $, $ H_b $, and $ H_{int} $ denote the system, bath, and interaction Hamiltonians, respectively. Setting $ \hbar = 1 $, the bosonic environment consists of a set of modes whose Hamiltonian is $ H_b = \sum_k \omega_k b_k^\dagger b_k $, where $ \omega_k $ is the frequency of the $ k-$th mode, and $ b_k $, $b_k^\dagger$ are the corresponding annihilation and creation operators satisfying $ [b_k, b_k^\dagger] = 1 $. The interaction Hamiltonian is taken as $H_{int} = \sum_k(g_k^* L^\dagger b_k + g_k L b_k^\dagger)$, where $ L $ is the Lindblad operator and $ g_k $ the complex coupling strength to mode $ k $. 
Within the quantum state diffusion \cite{PhysRevLett.83.4909} framework the system wave function is obtained by projecting the total state onto the Bargmann coherent state of the bath:
\begin{equation}
	\begin{aligned}
		\frac{\partial}{\partial t}\left|\psi(t,z_{t}^{*},w_{t}^{*})\right\rangle\:
		&=\:[-iH_{s}+Lz_{t}^{*}+L^{\dagger}w_{t}^{*}-L^{\dagger}\overline{O}(t,z_{t}^{*},w_{t}^{*})\\
		&-L\overline{Q}(t,z_{t}^{*},w_{t}^{*})]\left|\psi(t,z_{t}^{*},w_{t}^{*})\right\rangle,
	\end{aligned}
	\label{equ:a1}
\end{equation}
where $z_{t}^{*}$ and $w_{t}^{*}$ are complex Gaussian noise. The auxiliary operators $\overline{O}$ and $\overline{Q}$, which capture environmental memory effects, are defined as:
\begin{equation}
	\overline{O}(t,z_t^*,w_t^*)\:=\:\int_0^tds\alpha(t,s)O(t,s,z_t^*,w_t^*),
	\label{equ:a2}
\end{equation}
\begin{equation}
	\overline{Q}(t,z_t^*,w_t^*)\:=\:\int_0^tds\beta(t,s)Q(t,s,z_t^*,w_t^*),
	\label{equ:a3}
\end{equation}
with functional-derivative ansatz
\begin{equation}
	O(t,s,z_t^*,w_t^*)\left|\psi(t,z_t^*,w_t^*)\right\rangle=\frac{\delta}{\delta z_s^*}\left|\psi(t,z_t^*,w_t^*)\right\rangle,
	\label{equ:a4}
\end{equation}
\begin{equation}
	Q(t,s,z_t^*,w_t^*)\left|\psi(t,z_t^*,w_t^*)\right\rangle=\frac{\delta}{\delta w_s^*}\left|\psi(t,z_t^*,w_t^*)\right\rangle.
	\label{equ:a5}
\end{equation}
The operators \( O \) and \( Q \) encode how the state \( |\psi(t, z^*_t, w^*_t)\rangle \) at time \( t \) depends functionally on the earlier noise realizations \( z^*_s \) and \( w^*_s \) (with \( s \le t \)). The bath correlation function
\begin{equation}
	\alpha(t,s)=\int d\omega J(\omega)(\overline{n}_k+1)e^{-i\omega_k(t-s)},
	\label{equ:a6}
\end{equation}
\begin{equation}
	\beta(t,s)=\int d\omega J(\omega)\overline{n}_ke^{i\omega_k(t-s)}.
	\label{equ:a7}
\end{equation}
where \( \overline{n}_k = \bigl(e^{\omega_k/T} - 1\bigr)^{-1} \) is the average thermal occupation number of mode \(\omega_k\) and \( J(\omega) = \frac{\Gamma}{\pi} \frac{\omega}{1 + (\omega/\gamma)^2} \) is the bath spectral density \cite{ritschel2014analytic,WANG201078}. Consistency requires
\begin{equation}
	\begin{aligned}
		\frac{\partial O}{\partial t}=
		&\left[-\mathrm{i}H_{s}+Lz_{t}^{*}-L^{\dagger}\overline{O}+L^{\dagger}w_{t}^{*}-L\overline{Q},O\right]\\
		&-\left(L^{\dagger}\frac{\delta\overline{O}}{\delta z_{s}^{*}}+L\frac{\delta\overline{Q}}{\delta z_{s}^{*}}\right),
	\end{aligned}
	\label{equ:a8}
\end{equation}
\begin{equation}
	\begin{aligned}
		\frac{\partial Q}{\partial t}=
		&\left[-\mathrm{i}H_{s}+Lz_{t}^{*}-L^{\dagger}\overline{Q}+L^{\dagger}w_{t}^{*}-L\overline{Q},O\right]\\
		&-\left(L^{\dagger}\frac{\delta\overline{O}}{\delta z_{s}^{*}}+L\frac{\delta\overline{Q}}{\delta z_{s}^{*}}\right).
	\end{aligned}
	\label{equ:a9}
\end{equation}
The reduced density matrix is the ensemble average 
\begin{equation}
	\rho_s=M[P_t]\:=\:|\psi(t,z_{t}^{*},w_{t}^{*})\rangle\:\langle\psi(t,z_{t}^{*},w_{t}^{*})|],
	\label{equ:a10}
\end{equation}
with $ M[\cdot]=\prod_k\frac1\pi\int d^2ze^{-|z|^2}(\cdot)$.
Taking the time derivative gives the exact non-Markovian master equation \cite{PhysRevA.58.1699}
\begin{equation}
	\begin{aligned}
		\frac{\partial}{\partial t}\rho_{s}=
		&-i\left[H_{s},\rho_{s}\right]+[L,M[P_{t}\overline{O}^{\dagger}(t,z_{t}^{*},w_{t}^{*})]]\\
		&-[L^{\dagger},M[\overline{O}(t,z_{t}^{*},w_{t}^{*})P_{t}]]\\
		&+[L^{\dagger},M[P_{t}\overline{Q}^{\dagger}(t,z_{t}^{*},w_{t}^{*})]]\\
		&-[L,M[\overline{Q}(t,z_{t}^{*},w_{t}^{*})P_{t}]].
	\end{aligned}
	\label{equ:a11}
\end{equation}
The operators \(O\) and \(Q\) containing stochastic noise \(z_t^*\) and \(w_t^*\) are usually obtained approximately by perturbative techniques: for weak coupling, only the zeroth-order term is retained. Under this approximation, \(M[P_t \overline{O}^\dagger(t, z_t^*,w_t^*)] = \rho_s \overline{O}(t)\), \(M[P_t \overline{Q}^\dagger(t, z_t^*,w_t^*)] = \rho_s \overline{Q}(t)\) and Eq.~(1) reduces to the closed non-Markovian master equation quoted in the main text.

\section{Backpropagation and gradient flow in FPINN}
The backpropagation process of FPINN follows the chain rule, with gradients flowing precisely through the computational graph. For clarity, we define the network parameters as follows: the shared layer parameters \( \theta_s \), which include all weights and biases of the three shared fully connected layers. Then, the parameters of the \( O \) branch \( \theta_o \), which include the parameters of the middle layer \( \theta_o^{\mathrm{mid}} \) and the output layer \( \theta_o^{\mathrm{out}} \). Similarly, there are the \( Q \) branch parameters \( \theta_q \), which include the parameters of the middle layer \( \theta_q^{\mathrm{mid}} \) and the output layer \( \theta_q^{\mathrm{out}} \). For simplicity, the branches corresponding to operators \( \overline{O} \) or \( \overline{Q} \) are denoted as \( O \) and \( Q \) respectively.

FPINN takes time \(  t=(t_0,t_1,\ldots,t_f) \) as input. The forward propagation process is formalized into two parts: common feature extraction by the shared layers and unique feature separation by the branch layers. The shared feature extraction is expressed as
\begin{equation}
	h_1=\sigma(W_s^{(1)}t+b_s^{(1)}),
	\label{equ:b1}
\end{equation}
\begin{equation}
	h_2=\sigma(W_s^{(2)}h_1+b_s^{(2)}),
	\label{equ:b2}
\end{equation}
\begin{equation}
	h_3=\sigma(W_s^{(3)}h_2+b_s^{(3)}),
	\label{equ:b3}
\end{equation}
where \( \sigma \) is the activation function (SiLU in this work) and \( \theta_s = \{W_s^{(i)}, b_s^{(i)}\}_{i=1}^3 \). The branch-specific processing is given by
\begin{equation}
	O_{\mathrm{mid}}=\sigma(W_o^\mathrm{mid} h_3+b_o),\quad\theta_o^{\mathrm{mid}}=\{W_o^\mathrm{mid},b_o\},
	\label{equ:b4}
\end{equation}
\begin{equation}
	O_{\mathrm{out}}=W_o^\mathrm{out}O_\mathrm{mid}+b_o^\mathrm{out},\quad\theta_o^\mathrm{out}=\{W_o^\mathrm{out},b_o^\mathrm{out}\},
	\label{equ:b5}
\end{equation}
\begin{equation}
	Q_{\mathrm{mid}}=\sigma(W_q^\mathrm{mid} h_3+b_q),\quad\theta_q^\mathrm{mid}=\{W_q^\mathrm{mid},b_q\},
	\label{equ:b6}
\end{equation}
\begin{equation}
	Q_\mathrm{out}=W_q^\mathrm{out}Q_\mathrm{mid}+b_q^\mathrm{out},\quad\theta_q^\mathrm{out}=\{W_q^\mathrm{out},b_q^\mathrm{out}\},
	\label{equ:b7}
\end{equation}

During backpropagation, the total loss is
\begin{equation}
	\mathcal{L}_{tot}=\mathcal{L}_{o}(\theta_s,\theta_o)+\mathcal{L}_{q}(\theta_s,\theta_q).
	\label{equ:b8}
\end{equation}
According to the chain rule for multivariate functions, the gradient of the $ O-$branch parameters is: 
\begin{equation}
	\frac{\partial \mathcal{L}_{tot}}{\partial \theta_{o}} = \frac{\partial \mathcal{L}_{o}}{\partial \theta_{o}} + \frac{\partial \mathcal{L}_{q}}{\partial \theta_{o}}.
	\label{equ:b9}
\end{equation}
Since the calculation of \( \mathcal{L}_{q} \) does not depend on \( \theta_{o} \) (there is no connecting path in the computational graph),
\begin{equation}
	\frac{\partial \mathcal{L}_{q}}{\partial \theta_{o}} = 0 . 
	\label{equ:b10}
\end{equation}
and therefore
\begin{equation}
	\frac{\partial \mathcal{L}_{tot}}{\partial \theta_{o}} = \frac{\partial \mathcal{L}_{o}}{\partial \theta_{o}}.
	\label{equ:b11}
\end{equation}
Concretely, for the output layer weight,
\begin{equation}
	\frac{\partial \mathcal{L}_{o}}{\partial W_{o}^{\mathrm{out}}} = \frac{\partial \mathcal{L}_{o}}{\partial O_{\mathrm{out}}} \cdot \frac{\partial O_{\mathrm{out}}}{\partial W_{o}^{\mathrm{out}}} = \delta_{o} \cdot o_{\mathrm{mid}}^{T}, 
	\label{equ:b12}
\end{equation}
where \( \delta_o=\partial\mathcal{L}_o/\partial O_\mathrm{out} \) is the gradient of the $ O $ loss with respect to the output. Similarly, the gradient of the $ Q-$branch parameters is
\begin{equation}
	\frac{\partial \mathcal{L}_{tot}}{\partial \theta_{q}} = \frac{\partial \mathcal{L}_{q}}{\partial \theta_{q}}.
	\label{equ:b13}
\end{equation}
with the explicit form
\begin{equation}
	\frac{\partial \mathcal{L}_{q}}{\partial W_{q}^{\mathrm{out}}} = \delta_{q} \cdot q_{\mathrm{mid}}^{T},
	\label{equ:b14}
\end{equation}
where \( \delta_q=\partial\mathcal{L}_q/\partial Q_\mathrm{out} \).

The shared layer parameters are affected by both loss functions simultaneously:
\begin{equation}
	\frac{\partial\mathcal{L}_{{tot}}}{\partial\theta_{s}}=\frac{\partial\mathcal{L}_{o}}{\partial\theta_{s}}+\frac{\partial\mathcal{L}_{q}}{\partial\theta_{s}}.
	\label{equ:b15}
\end{equation}
Expanding the first term on the right-hand side of the equation (taking $  W_{s}^{(3)}  $ as an example) yields
\begin{equation}
	\frac{\partial\mathcal{L}_{o}}{\partial W_s^{(3)}}=\frac{\partial\mathcal{L}_{o}}{\partial O_{\mathrm{mid}}}\cdot\frac{\partial O_{\mathrm{mid}}}{\partial h_3}\cdot\frac{\partial h_3}{\partial W_s^{(3)}},
	\label{equ:b16}
\end{equation}
where 
\begin{equation}
	\frac{\partial\mathcal{L}_{o}}{\partial O_{\mathrm{mid}}}=\frac{\partial\mathcal{L}_{o}}{\partial O_{\mathrm{out}}}\cdot\frac{\partial O_{\mathrm{out}}}{\partial O_{\mathrm{mid}}}=\delta_{o}\cdot(W_{o}^{\mathrm{out}})^{T}.
	\label{equ:b17}
\end{equation}
Analogously, the second term is
\begin{equation}
	\frac{\partial\mathcal{L}_{q}}{\partial W_s^{(3)}}=\frac{\partial\mathcal{L}_{q}}{\partial Q_{\mathrm{mid}}}\cdot\frac{\partial Q_{\mathrm{mid}}}{\partial h_3}\cdot\frac{\partial h_3}{\partial W_s^{(3)}}
	\label{equ:b18}
\end{equation}
with
\begin{equation}
	\frac{\partial\mathcal{L}_{q}}{\partial Q_{\mathrm{mid}}}=\delta_q\cdot(W_q^\mathrm{out})^T.
	\label{equ:b19}
\end{equation}

The gradient flows for $ \overline{O} $ and $ \overline{Q} $ strictly follow the paths \( \delta_o \rightarrow \theta_o \rightarrow h_3 \rightarrow \theta_s \) and \( \delta_q \rightarrow \theta_q \rightarrow h_3 \rightarrow \theta_s \), respectively. FPINN employs PyTorch's automatic differentiation (AD) mechanism to implement gradient backpropagation \cite{paszke2017automatic}. When the backward pass is invoked, the computational graph is traversed backward from \( \mathcal{L}_{tot} \). Because the graph topology ensures no direct connections between \( \theta_o \) and \( \theta_q \), gradient propagation naturally achieves branch separation. The shared layer node \( h_3 \), as the common parent node of the two branches, accumulates gradients from the two subgraphs, with its total gradient being the sum of the gradients from the two independent backward paths. This design enables the shared layers to learn common features of \( \overline{O} \) and \( \overline{Q} \) (such as shared temporal evolution patterns), while the branched layers capture their unique dynamical characteristics. Consequently, the multi-objective optimization conflict between the two auxiliary operators is resolved, providing an optimized parameter learning mechanism for accurately modeling non-Markovian dynamics.

\section{The average Frobenius norm error and the average fidelity}
To quantify the accuracy of PINN for simulating the auxiliary operators $ \overline{O} $ and $ \overline{Q} $, we compute the average Frobenius-norm error relative to reference solutions obtained by a fourth-order Runge–Kutta (RK4) integration. Taking the \( \overline{O} \) operator as an example, the error is defined as
\begin{equation}
	\bar{\epsilon} = \frac{1}{f} \sum_{i=1}^{f} \left\| \overline{O}_{\text{PINN}}(t_i) - \overline{O}_{\text{RK4}}(t_i) \right\|_F,
	\label{equ:c1}
\end{equation}
where \( f \) is the total number of time steps, $ \overline{O}_{\text{PINN}}(t_i) $ and $ \overline{O}_{\text{RK4}}(t_i) $ are the \( 2^N \times 2^N \) complex matrices (\( N \) is the number of qubits) produced by the two methods at time $ t_k $. The Frobenius norm for a matrix $ A $ is
\begin{equation}
	\| A \|_F = \sqrt{\operatorname{tr}\left( A^\dagger A \right)} = \sqrt{\sum_{i=1}^{2^N} \sum_{j=1}^{2^N} |a_{ij}|^2}.
	\label{equ:c2}
\end{equation}
with \( |a_{ij}|^2 = [\operatorname{Re}(a_{ij})]^2 + [\operatorname{Im}(a_{ij})]^2 \) for each complex entry. This metric accumulates the element‑wise differences over the entire time evolution, preserving the full structural information of the complex matrices while avoiding cancellation of errors of opposite sign. A small value of \( \bar{\epsilon} \) indicates close agreement between PINN and the reference RK4 results, thereby validating the PINN approach for simulating non‑Markovian auxiliary operators.

For the reduced density matrix $ \rho(t) $, we assess the simulation quality via the average fidelity
\begin{equation}
	\bar F = \frac{1}{f} \sum_{i=1}^{f}
	\left[\operatorname{Tr}\sqrt{\sqrt{\rho_{\text{PINN}}(t_i)}\,\rho_{\text{RK4}}(t_i)\sqrt{\rho_{\text{PINN}}(t_i)}}\,\right]^2.
	\label{equ:c3}
\end{equation}
A value of $ \bar F $ near unity implies that the PINN‑generated trajectory closely follows the true quantum dynamics, confirming the method’s capability to accurately capture non‑Markovian system evolution.

\section{FPINN simulation of non-Markovian dynamics for the spin-boson model}

For the two level spin-boson model, the system Hamiltonian is \( H_s = \sigma_z \), the Lindblad operator is \( L = \sigma_x \), and the initial state of the system is \( \rho_0 = |0\rangle\langle0| \). The operators \( \overline{O} \) and \( \overline{Q} \) have vanishing diagonal elements; only their off‑diagonal entries are nontrivial. Consequently, the free real parameters describing these operators are \( N_o(t) = \{\text{Re}(\overline{O}_{12}), \text{Im}(\overline{O}_{12})\} \text{Re}(\overline{O}_{21}), \text{Im}(\overline{O}_{21})\} \) and analogously for $ N_q(t) $. For the density matrix, Hermiticity and unit trace reduce the independent variables to the triplet \( N_\rho(t) = \{\rho_{11}, \text{Re}(\rho_{12}), \text{Im}(\rho_{12})\} \). Thus, the PINN output for \(\overline{O}\) (or \(\overline{Q}\)) consists of four real numbers, while the output for $ \rho $ consists of three.

Fig.~\ref{fig:d1} displays the time evolution of the real and imaginary parts of \( \overline{O} \) and \( \overline{Q} \) simulated by FPINN for three values of the character frequency $ \gamma = 0.3, 0.5, 1 $ (with $ \Gamma = 0.1 $ and $ T = 20 $). The RK4 reference solutions are shown for comparison. The FPINN trajectories closely follow the RK4 results in all cases. To demonstrate the effect of the evolution regularization loss \( \mathcal{L}_{er} \), we trained three additional FPINN models using the exact same setup but without including the evolution regularization loss \( \mathcal{L}_{\text{er}} \) during training. The corresponding average Frobenius norm errors \( \bar{\epsilon} \) (defined in Eq.~(\ref{equ:c1})) are listed in Table~\ref{tab:1}. For models trained with the evolution regularization loss $ \mathcal{L}_{er} $, the errors remain below $ 1\% $ across all $ \gamma $ values, confirming the high accuracy of the FPINN simulations. Notably, when $ \mathcal{L}_{er} $ is omitted, the errors are substantially larger for small $ \gamma $ (strong non‑Markovian regime), highlighting the regularizer’s role in preventing premature stagnation.

Fig.~\ref{fig:d2} shows the changes in the loss functions during the training process of the models. Panels (b), (d), and (f) of Fig.~\ref{fig:d2} show the changes in $ \mathcal{L}_{ini} $ during the training process of the three models. After the models converge, $ \mathcal{L}_{ini} $ is several orders of magnitude smaller than the other loss terms and has a smaller impact on \( L_{\text{tot}} \). Panels (a), (c), and (e) of Fig.~\ref{fig:d2} show the changes in the remaining loss terms during the training process of the six models. It can be seen that whether trained with or without $ \mathcal{L}_{er} $, the models eventually converge to similar $ \mathcal{L}_{tot} $ values. However, in Fig.~\ref{fig:d2}(a), the models trained with $ \mathcal{L}_{er} $ have smaller $ \mathcal{L}_{mod}^o $ and $ \mathcal{L}_{mod}^q $ compared to those trained without $ \mathcal{L}_{er} $. This indicates that the models trained with $ \mathcal{L}_{er} $ produce \( \overline{O}(t) \) and \( \overline{Q}(t) \) that more closely satisfy their constraint equations, implying more accurate simulations. As \( \gamma \) increases, this advantage gradually diminishes, which corresponds to the results in Fig.~\ref{fig:d1} and further confirms the role of $ \mathcal{L}_{er} $ in simulating non-Markovian dynamics with FPINN.

Using the simulated \( \overline{O}(t) \) and \( \overline{Q}(t) \) as prior knowledge, we then compute the reduced density matrix \( \rho(t) \). The evolution of the observable \( \langle\sigma_z\rangle(t)=\mathrm{Tr}[\rho(t)\sigma_z] \) is shown in Fig.~\ref{fig:d3}(a). The FPINN results again agree excellently with the RK4 curves for all three $ \gamma $ values, yielding average fidelities $ \bar F $ (Eq.~\eqref{equ:c3}) of $ 0.999951 $, $ 0.999976 $, and $ 0.999986 $, respectively. As $ \gamma $ increases, the environment approaches the Markovian limit, and the characteristic backflow of \( \langle \sigma_z \rangle \) gradually diminishes—a trend faithfully reproduced by FPINN. This consistency further validates the reliability of FPINN for simulating non‑Markovian quantum dynamics across a range of memory strengths. Fig.~\ref{fig:d3}(b) shows the convergence of the total loss function $ \mathcal{L}_{tot} $ of the models during the training process.

\begin{table}[htbp]
	\centering
	\caption{The average Frobenius norm error \( \bar{\epsilon} \) for the operators \( \overline{O} \) and \( \overline{Q} \) obtained with and without the evolution regularization loss $ \mathcal{L}_{er} $.}
	\begin{tabular*}{\columnwidth}{@{\extracolsep{\fill}}cccc}
		\hline
		\( \bar{\epsilon} \)			                 & $\gamma=0.3 $    & $\gamma=0.5$  & $\gamma=1$	\\
		\hline
		$\overline{O}$ (with $ \mathcal{L}_{er} $)	     & 0.002914 		& 0.005891      & 0.004134   	 \\
		$\overline{Q}$ (with $ \mathcal{L}_{er} $ )      & 0.003061	    	& 0.005953 	    & 0.003915  	 \\
		$\overline{O}$ (without $ \mathcal{L}_{er} $)    & 0.029402			& 0.022747 	    & 0.002381	 	 \\
		$\overline{Q}$ (without $ \mathcal{L}_{er} $)	 & 0.029932    		& 0.022693 	    & 0.008990	     \\
		\hline
	\end{tabular*}
	\label{tab:1}
\end{table}	

\begin{figure*}  
	\centering
	\includegraphics[width=0.95\textwidth]{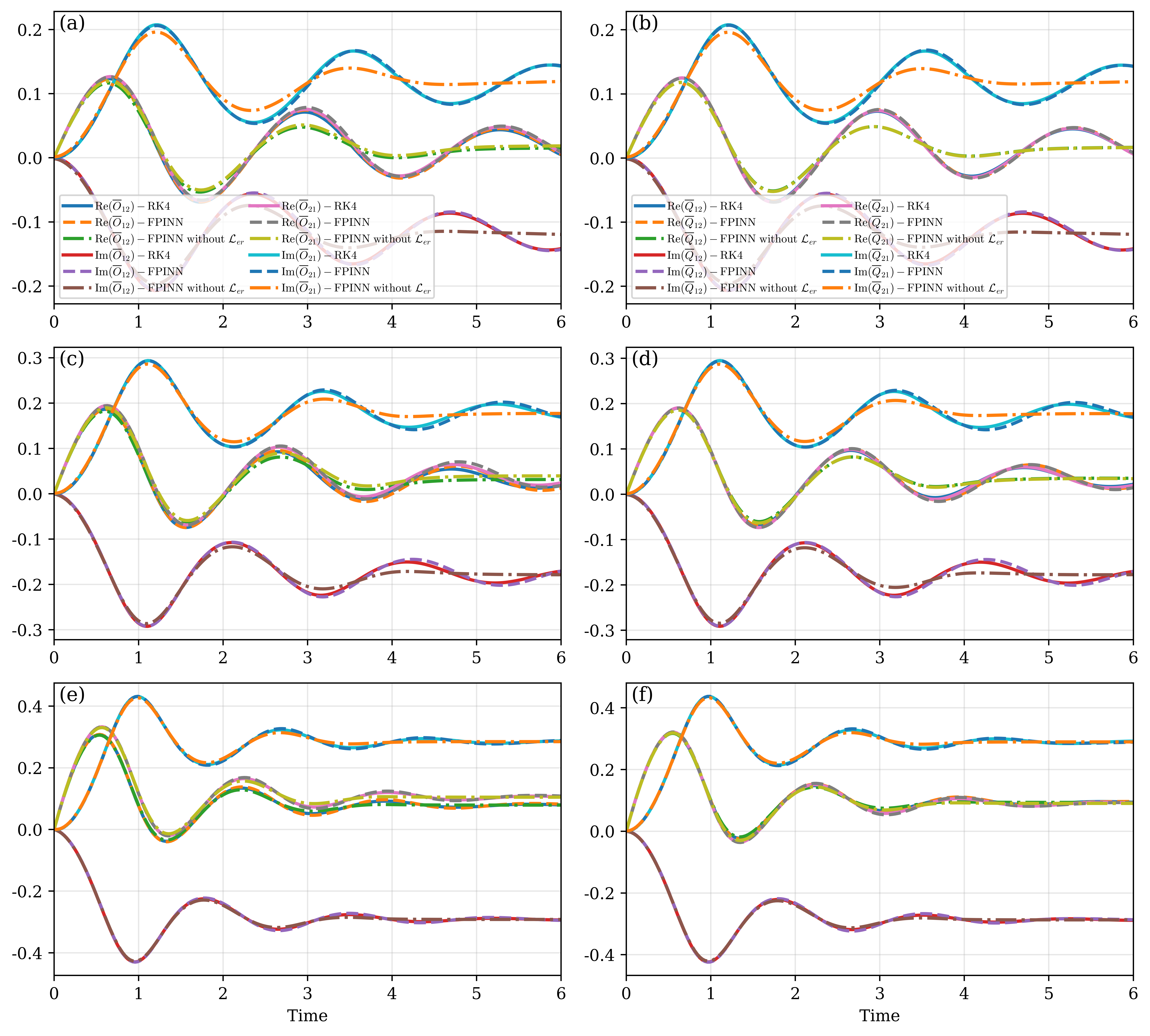}
	\caption{Evolution of the operators \( \overline{O} \) and \( \overline{Q} \) for the spin‑boson model simulated by FPINN and RK4 for different $ \gamma $. Panels (a,c,e) show \( \overline{O}(t) \), panels (b,d,f) show \( \overline{Q}(t) \). The first row has parameters \( \gamma = 0.3 \), the second row \( \gamma = 0.5 \), and the third row \( \gamma = 0.7 \) (\( \Gamma = 0.1 \) and \( T = 20 \)). The sampling time points \( t_f = 201 \), the maximum evolution time \( T_{\text{tot}} = 6 \).} 
	\label{fig:d1}
\end{figure*}

\begin{figure*}  
	\centering
	\includegraphics[width=0.95\textwidth]{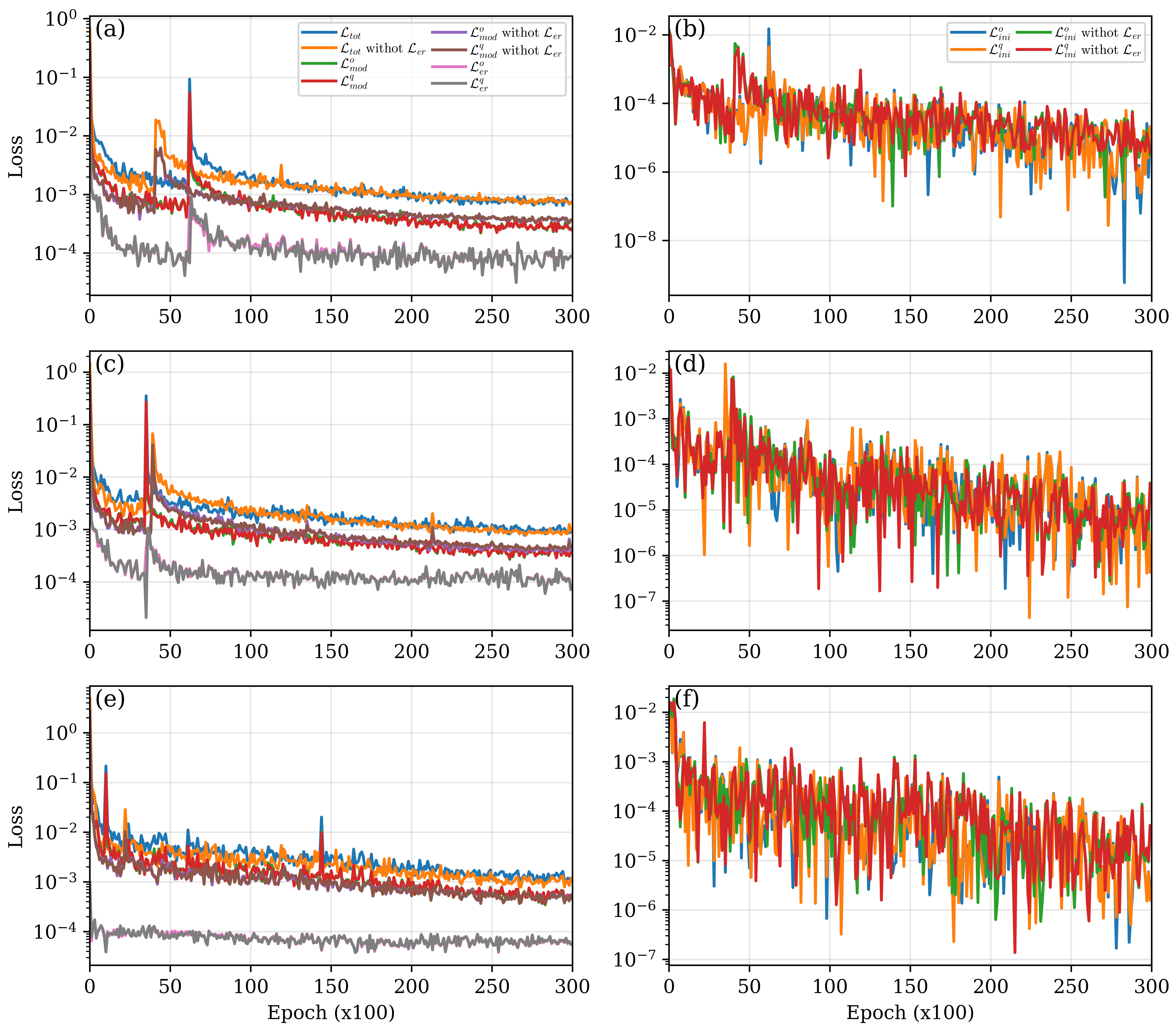}
	\caption{The change of the loss function during the training process of the FPINN model with and without \( L_{\text{er}} \). Panels (a, c, e) show \( \mathcal{L}_{\text{tot}} \), \( \mathcal{L}_{\text{mod}} \), and \( \mathcal{L}_{\text{er}} \), while panels (b, d, f) show \( \mathcal{L}_{\text{ini}} \). The first row has parameters \( \gamma = 0.3 \), the second row \( \gamma = 0.5 \), and the third row \( \gamma = 0.7 \) (\( \Gamma = 0.1 \) and \( T = 20 \)).} 
	\label{fig:d2}
\end{figure*}

\begin{figure}[htbp] 
	\centering{\includegraphics[width=\columnwidth]{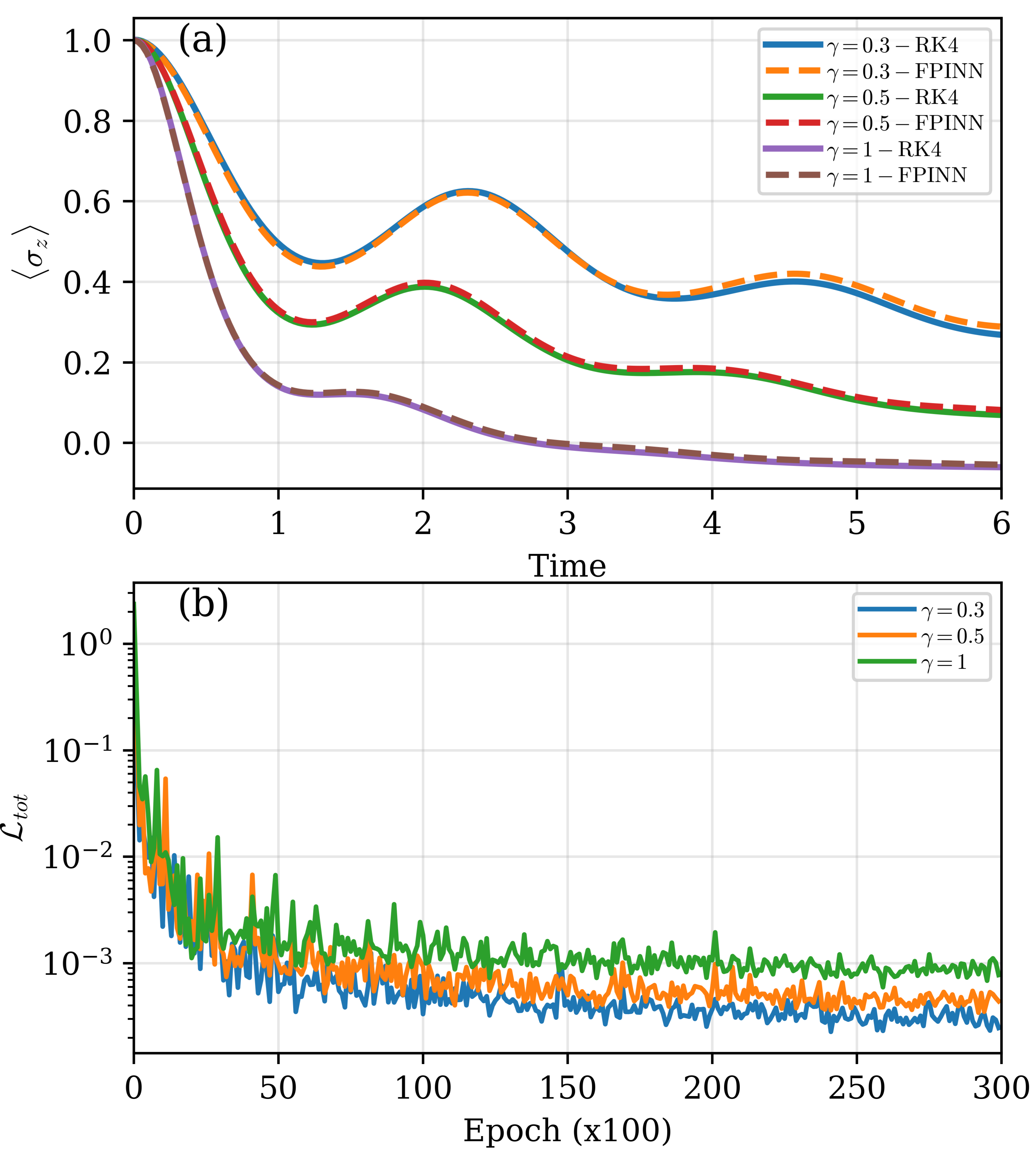}}
	\caption{(a) Evolution of \( \langle \sigma_z \rangle \) for the spin‑boson model under different \( \gamma \) values, comparing FPINN simulations with RK4 reference solutions. (b) Comparison of the convergence of the loss function during the training process of FPINN trained under different \( \gamma \) values.} 
	\label{fig:d3}
\end{figure}

\section{Unified PINN (UPINN) and Separated PINN (SPINN)}	

To systematically assess the advantages of FPINN for coupled physical systems, we compare it with two alternative architectures: the Unified PINN (UPINN) and the Separated PINN (SPINN).

UPINN (Fig.~\ref{fig:e1}(a)) employs a single ``straight‑tube” network. The time variable \( t \) is passed through a sequence of fully connected hidden layers to produce a shared feature representation. At the output layer, two separate heads generate the prediction vectors \( N_o(t) \) and \( N_q(t) \) for the operators  \(\overline{O} \) and \( \overline{Q} \). UPINN is optimized with the same total loss \( \mathcal{L}_{tot} \) as FPINN. During backpropagation, however, the gradients from the two operator losses are summed and applied uniformly to update all hidden‑layer parameters. While mathematically valid, this gradient‑summation scheme can lead to conflicting directions when the dynamical behaviors of \( \overline{O} \) and \( \overline{Q} \) differ substantially, thereby hindering convergence and accuracy.

SPINN (Fig.~\ref{fig:e1}(b)) adopts a fully decoupled dual network design. Two independent subnetworks, each with identical architecture (input layer, three hidden layers, output layer) but separate parameters, are dedicated to predicting \(\overline{O} \) and \( \overline{Q} \), respectively. Both networks receive the same time input \( t \) and output their corresponding prediction vectors \( N_o(t) \) and \( N_q(t) \). The predictions are pooled to compute the total loss \( \mathcal{L}_{tot} \). In the backward pass, the gradient of \( \mathcal{L}_{tot} \) with respect to \( \overline{O} \) updates only the first subnetwork, and the gradient with respect to \( \overline{Q} \) updates only the second. No gradient information is shared between the two subnetworks. Although this separation avoids gradient conflict, it ignores the intrinsic coupling between \( \overline{O} \) and \( \overline{Q} \) encoded in the physical equations. Consequently, SPINN cannot exploit co‑evolutionary information, and the independent optimization may lead to asynchronous convergence, where one subnetwork stagnates prematurely while the other is still training, ultimately degrading overall prediction fidelity.

Taking the two level spin-boson model (\( \gamma = 0.3 \)) as an example, we plot the changes in the total loss function during the training process of the three network structures in Fig.~\ref{fig:e2}. Under the same number of neurons and the same training method, FPINN, with its superior network structure, achieves a total loss that is nearly an order of magnitude smaller than that of the other two networks after convergence. This indicates that FPINN has a greater advantage in dealing with systems of nonlinear partial differential equations.

\begin{figure}[htbp] 
	\centering{\includegraphics[width=\columnwidth]{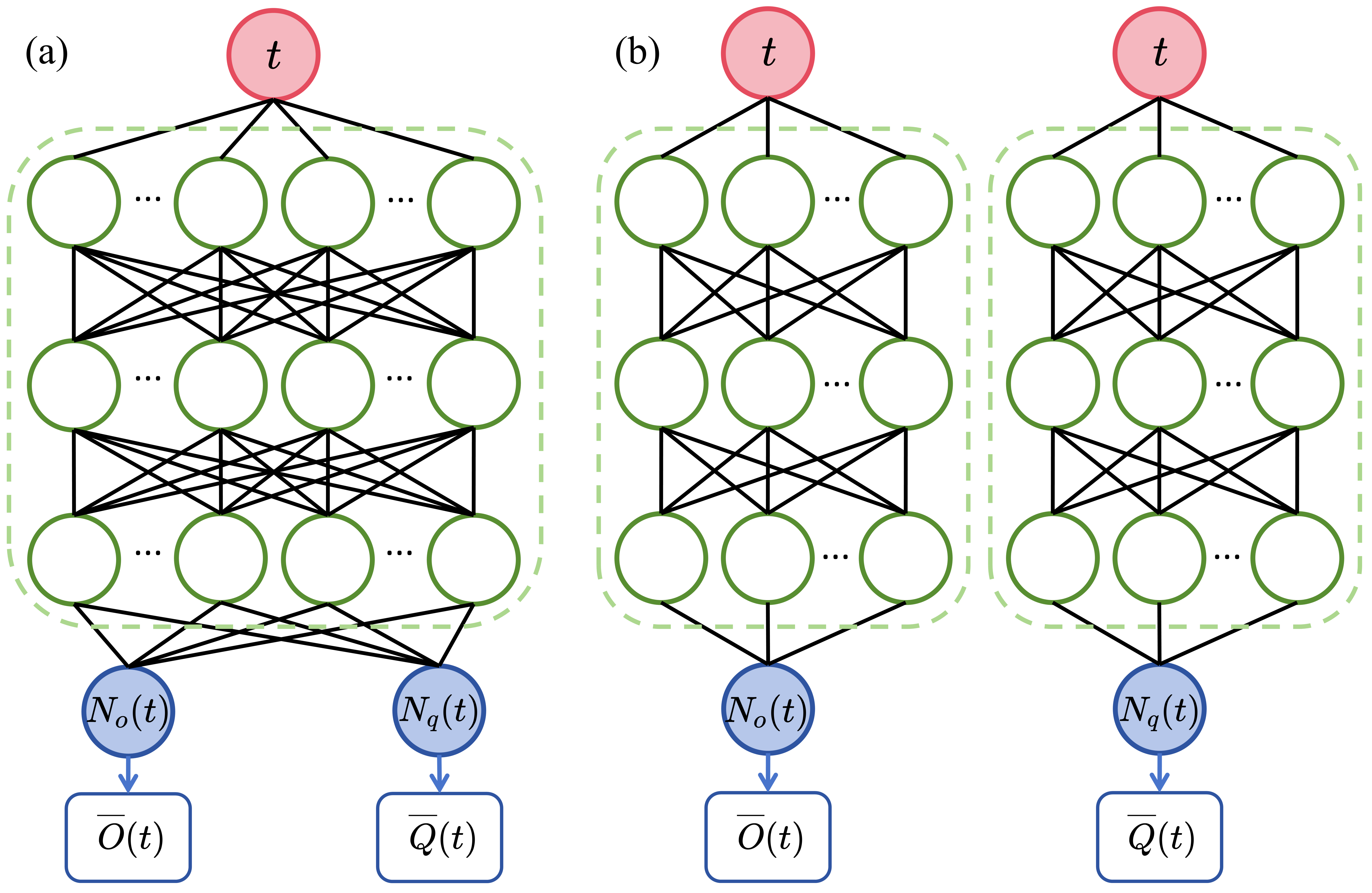}}
	\caption{Schematic of the UPINN and SPINN architectures. (a) UPINN uses a single shared hidden‑layer stack; separation occurs only at the output layer, which produces \(N_o(t)\) and \(N_q(t)\). (b) SPINN consists of two isolated networks of identical structure, each dedicated to outputting either \(N_o(t)\) or \(N_q(t)\). For fair comparison, the number of neurons in one hidden layer of each SPINN subnetwork is half that of the corresponding layer in UPINN.} 
	\label{fig:e1}
\end{figure}

\begin{figure}[htbp] 
	\centering{\includegraphics[width=\columnwidth]{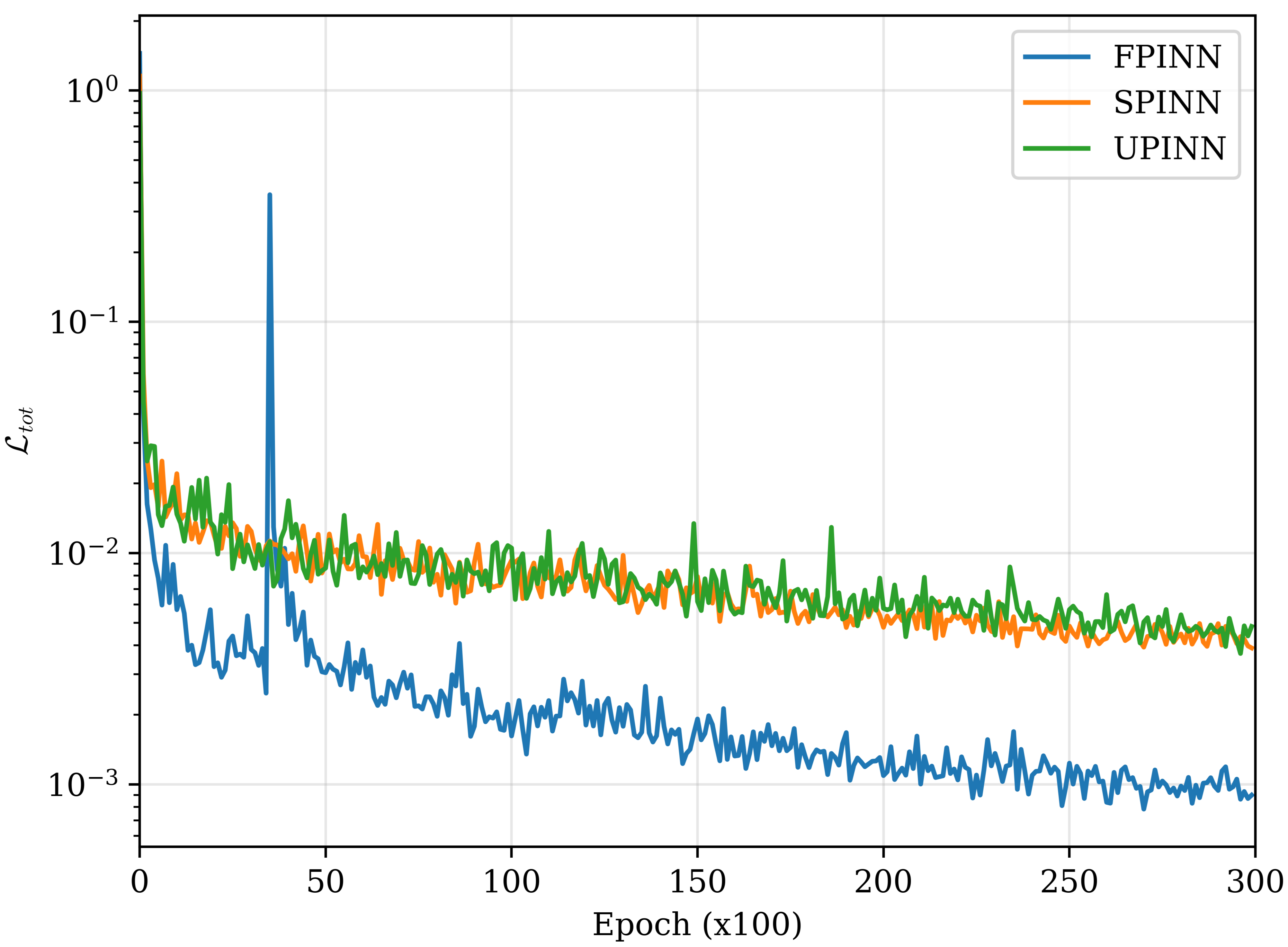}}
	\caption{Evolution of the total loss function during the training process of the three network architectures.} 
	\label{fig:e2}
\end{figure}

\section{FPINN Simulation of Non-Markovian Dynamics for the XXZ Model}

Moving to the two-qubit Heisenberg XXZ chain, \( H = J(\sigma_x^1\sigma_x^2 + \sigma_y^1\sigma_y^2) + \Delta(\sigma_z^1\sigma_z^2) \), we set $ J =2 $ and $ \Delta=0.5$. Under this model, the auxiliary operator \( \overline{O} \) satisfies \( \overline{O}_{21} = \overline{O}_{31} \), \( \overline{O}_{42} = \overline{O}_{43} \), with all other matrix elements being zero; the auxiliary operator \( \overline{Q} \) satisfies \( \overline{O}_{12} = \overline{O}_{13} \), \( \overline{O}_{24} = \overline{O}_{34} \), with all other matrix elements being zero. Therefore, the feature vectors output by FPINN are \( N_o(t) = \{\text{Re}(\overline{O}_{21}), \text{Im}(\overline{O}_{21}), \text{Re}(\overline{O}_{42}), \text{Im}(\overline{O}_{42})\} \) and \( N_q(t) = \{\text{Re}(\overline{O}_{12}), \text{Im}(\overline{O}_{12}), \text{Re}(\overline{O}_{24}), \text{Im}(\overline{O}_{24})\} \). Fig.~\ref{fig:f1} shows the convergence of the loss function during the training process of the FPINN simulating the dynamics of the operators \( \overline{O} \) and \( \overline{Q} \) for the XXZ model. 

The evolution of the operators \( \overline{O} \) and \( \overline{Q} \) are independent of the system's initial state. Thus, for the same environment and quantum system, \( O(t) \) and \( Q(t) \) can be used to help calculate the evolution of different initial states. We used the simulated \( O(t) \) and \( Q(t) \) as prior information to train two PINN models for simulating the dynamical evolution of two different initial states: \( |00\rangle \) and the Bell state \( \frac{1}{\sqrt{2}}(|00\rangle + |11\rangle) \). Fig.~\ref{fig:f2} shows the changes in the loss function during the training process of these two models.

\begin{figure}[htbp] 
	\centering{\includegraphics[width=\columnwidth]{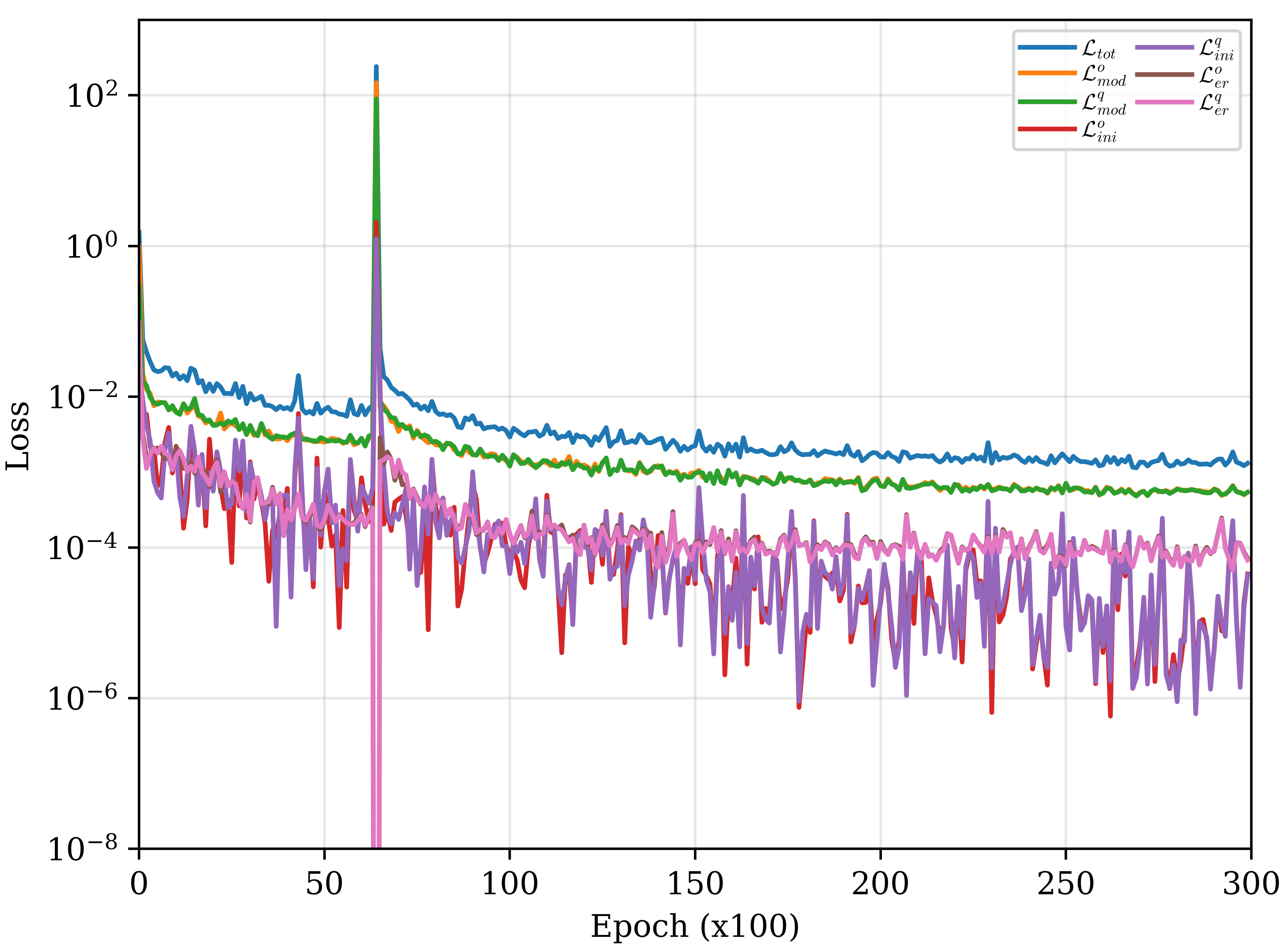}}
	\caption{The change of the loss function during the training process of the FPINN model for simulating the auxiliary operator \( \overline{O}(t) \) and \( \overline{Q}(t) \) of the XXZ model.} 
	\label{fig:f1}
\end{figure}

\begin{figure}[htbp] 
	\centering{\includegraphics[width=\columnwidth]{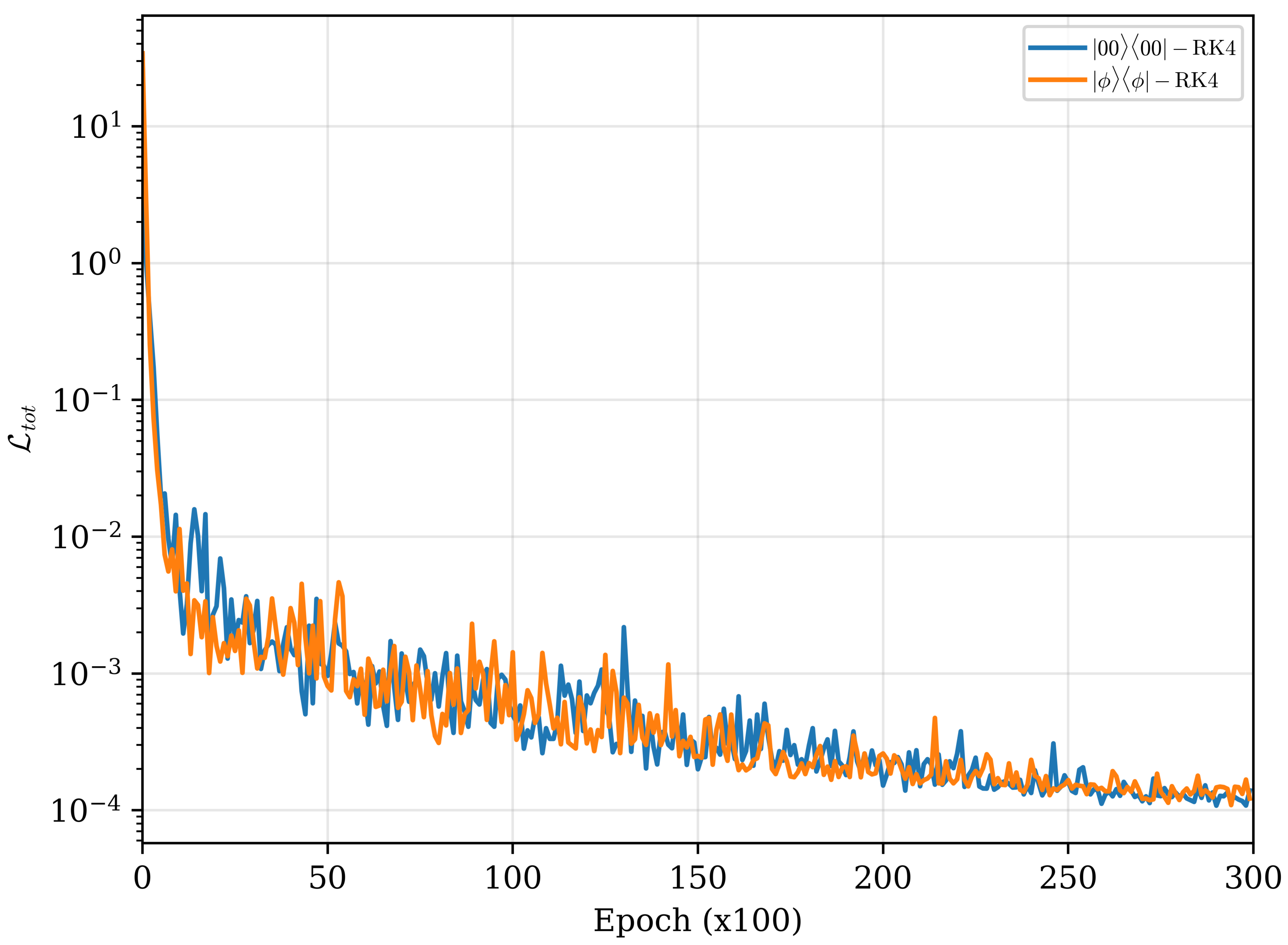}}
	\caption{The change of the loss function during the training process of the FPINN model for simulating the density matrix \( \rho \)  of the XXZ model.} 
	\label{fig:f2}
\end{figure}

\section{Summary of PINNs parameters}

\begin{table*}[t]
	\centering
	\caption{Hyperparameter information of different PINN models.}
	\begin{tabular*}{\textwidth}{@{\extracolsep{\fill}}ccccc}
		\hline
		Simulation task	        & Spin-Boson Model & Spin-Boson Model & Spin-Boson Model & XXZ Model	\\
		\hline
		Network Architecture	& FPINN & UPINN & SPINN & FPINN	\\
		Number of shared layers & 3 & 4 & 0 & 3	\\
		Number of separated layers & 1 & 0 & 4 & 1	\\
		Neurons per hidden layer  & 256 or 128,128 	& 256  & 128,128 & 256 or 128,128 	\\
		Initial learning rate $\eta_{0} $ & $ 5\times10^{-3} $ & $ 5\times10^{-3} $ & $ 5\times10^{-3} $ & $ 5\times10^{-3} $	\\
		Minimum learning rate $\eta_{\min}$ & $ 1\times10^{-5} $ & $ 1\times10^{-5} $ & $ 1\times10^{-5} $ & $ 1\times10^{-5} $    \\
		Total training epoch $ T_{max} $	& $ 3\times10^4 $ & $ 3\times10^4 $ & $ 3\times10^4 $ & $ 3\times10^4 $   \\
		Number of time samples $ t_f $	& 201 & 201 & 201 & 401   \\
		Total evolution time $ T_tot $ & 6 & 6 & 6 & 6   \\
		Evolution regularization loss weight $ \lambda_{er} $ & 0.01,0.01,0.001 & 0.01 & 0.01 & 0.1   \\
		
		\hline
	\end{tabular*}
	\label{tab:2}
\end{table*}	

To summarize the results of PINNs in simulating non-Markovian dynamics, this section presents the details of the hyperparameters and training strategies used during model training. In Table~\ref{tab:2}, we present the relevant hyperparameter information during the training process of different models. Each fully connected layer in PINNs is followed by a SiLU activation function, Dropout regularization (dropout rate = 0.1), and layer normalization to enhance the model's nonlinear expression capability and training stability. The training of the model uses the AdamW optimizer \cite{loshchilov2017decoupled}, which decouples weight decay regularization from the gradient update step based on the standard Adam algorithm. The optimizer's configuration parameters are set as follows: initial learning rate \( \eta_0 = 5 \times 10^{-3} \), weight decay coefficient \( \lambda = 10^{-5} \), momentum parameters \( \beta_1 = 0.9 \) and \( \beta_2 = 0.999 \). This configuration, which adaptively adjusts the learning rates of individual parameters and explicitly constrains network weights through L2 regularization, helps improve the numerical stability of PINNs when solving partial differential equations.

To balance the convergence speed and final accuracy of the training process, a cosine annealing learning rate scheduler \cite{loshchilov2016sgdr} is introduced. Within the preset total training epoch \( T_{\text{max}} \) (i.e., the total number of iterations), the learning rate is smoothly reduced according to a cosine curve. The update formula is 
\begin{equation}
	\eta_t = \eta_{\min} + \frac{1}{2}(\eta_0 - \eta_{\min})\left(1 + \cos\left(\frac{T_{\text{cur}}}{T_{\text{max}}}\pi\right)\right), 
	\label{equ:g1}
\end{equation}
where \( \eta_{\min} = 10^{-5} \) is the minimum learning rate, and \( T_{\text{cur}} \) is the current training iteration. This dynamic scheduling mechanism maintains a high learning rate in the early training phase to accelerate convergence, then gradually decays to a lower level, facilitating fine-tuning of model parameters in the later phase. Overall, it balances optimization efficiency and convergence stability.


\end{document}